\documentclass[11pt]{article}       
\usepackage{geometry}               
\usepackage{graphicx}
\usepackage{caption}
\usepackage{subcaption}
\usepackage{enumerate}
\usepackage{hyperref}
\usepackage{cite}

\usepackage{amsfonts}
\usepackage{epstopdf}
\usepackage{caption}
\usepackage{amsmath,amsfonts,mathrsfs,amssymb,amscd}
\usepackage{amsthm}

\usepackage{mathtools}
\usepackage{bm}
\usepackage{lipsum}
\usepackage[linesnumbered,ruled,vlined]{algorithm2e}
\usepackage{color}
\usepackage{array}
\usepackage{multirow}
\usepackage{cite}
\usepackage{url}
\usepackage{booktabs}

\newtheorem{proposition}{Proposition}

\newtheorem{assumption}{Assumption}
\newtheorem{lemma}{Lemma}
\newtheorem{theorem}{Theorem}
\newtheorem{remark}{Remark}

\newtheorem{cor}{Corollary}

\DeclareMathOperator*{\argmin}{arg\,min}

\newcommand{\bbar}[1]{\bar{\bar{#1}}}

\title{Wasserstein Distributionally Robust Control of\\ Partially Observable Linear Stochastic Systems\thanks{This work was supported in part by  the National Research Foundation of Korea funded by  MSIT(2020R1C1C1009766, 2021R1A4A2001824), the Information and Communications Technology Planning and Evaluation  grant funded by MSIT(2022-0-00480), and Samsung Electronics. 
This paper is significantly extended from its preliminary version~\cite{Hakobyan2022conf}, where we introduced the WDRC method for the finite-horizon case. Specifically, we provide a nontrivial extension to the infinite-horizon average-cost setting and analyze new salient theoretical features, such as the out-of-sample performance guarantee and the stability of the closed-loop system.} }

\author{Astghik Hakobyan \and
 Insoon Yang\thanks{A. Hakobyan, and I. Yang are with the Department of Electrical and Computer Engineering and ASRI, Seoul National University, Seoul, 08826, South Korea {\tt\small \{astghikhakobyan, insoonyang\}@snu.ac.kr}}
}

\date{}

\begin{document}
\maketitle

\pagestyle{myheadings}
\thispagestyle{plain}

\begin{abstract}
Distributionally robust control (DRC) aims to effectively manage distributional ambiguity in stochastic systems. While most existing works address  inaccurate distributional information in fully observable settings, we consider a partially observable DRC problem for discrete-time linear systems using the Wasserstein metric.
 For a tractable solution, we propose a novel approximation method exploiting the Gelbrich bound of the Wasserstein distance.
 Using techniques from modern distributionally robust optimization, 
we derive a closed-form expression for the optimal control policy and a tractable semidefinite programming problem for the worst-case distribution policy in both finite-horizon and infinite-horizon average-cost settings.
 The proposed method features several salient theoretical properties, such as a guaranteed cost property and a probabilistic out-of-sample performance guarantee, demonstrating the distributional robustness of our controller. 
 Furthermore, the resulting controller is shown to ensure the closed-loop stability of the mean-state system. The empirical performance of our method is tested through numerical experiments on a power system frequency control problem.
\end{abstract}

\section{Introduction}\label{sec:intro}

Optimal control of linear dynamical systems under uncertainties has a long history and is regarded as one of the most fundamental topics in control theory~\cite{aastrom2012introduction}. In various practical systems, the system states are not entirely observable, and there is only partial information available about the system coming from the noisy measurements. The theory of optimal control handles such imperfect state information either in stochastic or robust control frameworks.
Robust optimal control methods address uncertainties in a pre-specified disturbance set and seek to find a controller concerning the worst-case realization of the disturbance~(e.g.,~\cite{khalil1996robust}).
However, the resulting controllers are often conservative as
 no   information other than the support of disturbances is used, and potentially useful statistical properties of the disturbances are disregarded.
On the contrary, stochastic optimal control approaches design a controller using the knowledge of the disturbance distribution, which is typically modeled as Gaussian~(e.g.,\cite{Kumar2015}). 
However, it is often difficult to obtain an accurate probability distribution of disturbances.
Using imperfect distributional information does not guarantee the optimality of the resulting controller and may even cause undesirable system behaviors~(e.g.,\cite{Nilim2005, Samuelson2017}).

To alleviate the aforementioned issues and bridge the gap between the two methods, 
distributionally robust control (DRC) has emerged as an alternative tool, balancing the tradeoff between required information and conservativeness~\cite{
petersen2000minimax,
ugrinovskii2002minimax,
van2015distributionally, 
Yang2018,
Tzortzis2019,
Coppens2021, schuurmans2021data, yang2020wasserstein, coulson2021distributionally, Mark2021, 
Tzortzis2021, hakobyan2021wasserstein, 
Zolanvari2021, kim2022minimax, Zhong2022, Dixit2022, Micheli2022}.
With DRC, a controller is designed to  minimize the expected cost of interest with respect to the worst-case probability distribution of disturbances in a so-called \emph{ambiguity set}. 
Thus, the resulting controller proactively manages possible deviations of the true distribution from the nominal one used in the controller design.

\begin{figure}[t]
    	\centering
    	\includegraphics[width=0.6\linewidth]{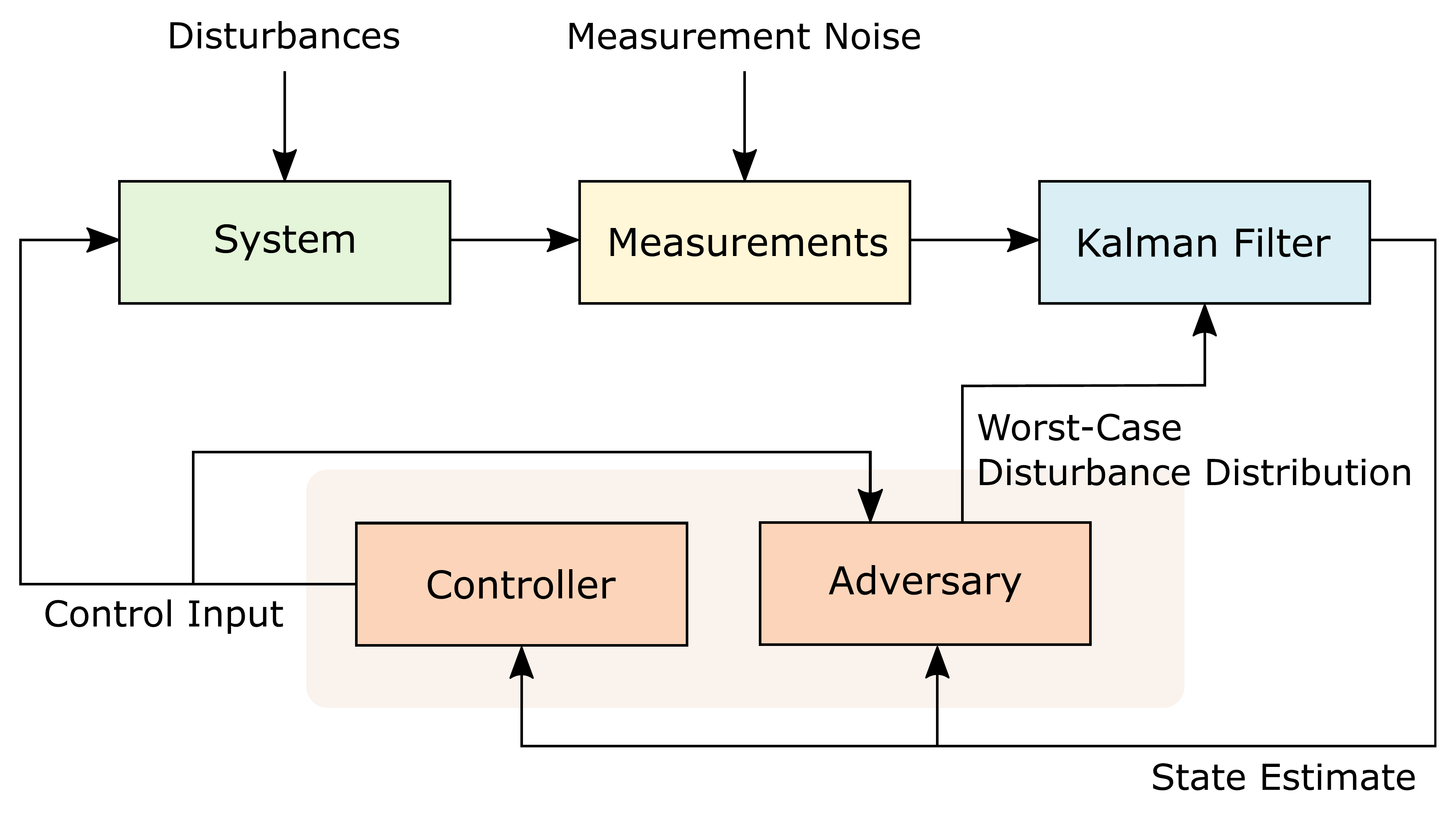}
    \caption{Block diagram of the proposed WDRC scheme.}
    \label{fig:overview}
\end{figure}

DRC can be regarded as a dynamic or multi-stage version of distributionally robust optimization (DRO).
In the literature regarding DRO, it is common to design the ambiguity set based on a nominal distribution constructed from data so that it contains the true distribution with high probability. For example, moment-based ambiguity sets are popular in DRO, which include distributions satisfying some moment constraints~\cite{calafiore2007ambiguous, delage2010distributionally, wiesemann2014distributionally}. Despite outstanding tractability properties, such sets often yield conservative decisions and require accurate moment estimates. Designing the ambiguity set based on statistical distances to contain distributions close to the given nominal one is another popular option. 
Among various distances, such as the KL-divergence and Prokhorov metric~\cite{ben2013robust, bayraksan2015data}, the Wasserstein metric attracts  significant attention not only in DRO~\cite{mohajerin2018data, gao2016distributionally, zhao2018data, kuhn2019wasserstein} but also in DRC~\cite{coppens2020data, Mark2020, yang2020wasserstein, coulson2021distributionally,kim2022minimax, hakobyan2021wasserstein}.  The Wasserstein ambiguity set has a number of useful features, including  offering a powerful finite-sample performance guarantee~\cite{mohajerin2018data,Boskos2020}.
Furthermore, it is rich enough to contain relevant distributions, thereby encouraging the DRO problem to avoid  providing pathological solutions~\cite{gao2016distributionally}.

In contrast to research on fully observable settings, the literature about partially observable DRC  is relatively sparse. A few works are devoted to the distributionally robust version of the linear-quadratic-Gaussian (LQG) control method. For example,~\cite{ugrinovskii1999finite, petersen2000minimax, ugrinovskii2002minimax} propose a minimax LQG controller that minimizes the worst-case performance by restricting the KL-divergence between the  disturbance distribution and a given reference distribution. In~\cite{nakao2021distributionally}, a partially observable Markov decision process is considered with finite state, action, and observation spaces. The ambiguity set is chosen to bound the moments of the joint distribution of the transition-observation probabilities. Another type of partially observable systems, namely the Markov jump linear system, is studied in~\cite{schuurmans2021data}. The authors propose a mechanism for estimating the active mode in a receding horizon fashion and integrate this procedure with a data-driven distributionally robust controller design using the total variation distance. 
In~\cite{coulson2021distributionally}, a data-enabled distributionally robust predictive control method is proposed and studied using noise-corrupted input and output data.

Departing from the existing literature,
our particular interest is in the Wasserstein DRC (WDRC) methods for partially observable linear-quadratic optimal control in discrete time, motivated by the superior properties of Wasserstein DRO. 
The WDRC problem is challenging to solve due to partial observability in addition to the infinite-dimensionality of the Wasserstein DRO problem in the Bellman equation. 
To resolve these issues, 
we propose a novel approximation technique for partially observable WDRC problems by replacing the Wasserstein ambiguity set with a special penalty term using the Gelbrich bound. 
The approximate problem is first solved in the finite-horizon setting by deriving a non-trivial Riccati equation alongside a closed-form expression for the optimal control policy. 
Then, we examine the asymptotic behavior of the controller and extend the results to the infinite-horizon average-cost setting. 
Consequently, we obtain optimal control and distribution policies by solving an algebraic Riccati equation (ARE) and a tractable semidefinite programming (SDP) problem. 
The overall scheme of the proposed WDRC method is illustrated in Fig.~\ref{fig:overview}. 

The proposed controller possesses several salient theoretical properties. First, it is shown to enjoy a guaranteed cost property for any worst-case disturbance distribution in the Wasserstein ambiguity set. This demonstrates the distributional robustness of our controller despite being constructed by solving an approximate WDRC problem. 
Second, the proposed controller offers a probabilistic out-of-sample performance guarantee. 
Last but not least,  the proposed controller is shown to ensure the stability of the closed-loop mean-state system as well as its bounded-input, bounded-output (BIBO) stability when viewing the disturbances as input.

The rest of this article is organized as follows. In Section~\ref{sec:prelim}, we introduce  the partially observable WDRC problem 
for linear systems. In Section~\ref{sec:tract_approx}, we introduce the tractable approximation and derive its solution in both finite- and infinite-horizon average-cost settings. In addition, we analyze the optimality of the resulting solution and describe the overall WDRC algorithm. In Section~\ref{sec:perf}, we present the guaranteed cost property and out-of-sample performance guarantee of our controller. Section~\ref{sec:stab} concerns the stability properties of the closed-loop mean-state system. Finally, Section~\ref{sec:exp} demonstrates the performance and utility of the proposed method through 
numerical experiments on a power system frequency control problem.

\section{Preliminaries}\label{sec:prelim}

\subsection{Notation}
We let $\mathcal{P}(\mathcal{W})$ denote the set of Borel probability measures with support $\mathcal{W}$. The expected value of function $f(x)$, where $x$ is a random variable with a probability distribution $\mathbb{P}$, is denoted by $\mathbb{E}_x[f(x)]$. We denote the space of all symmetric matrices in $\mathbb{R}^{n\times n}$ by $\mathbb{S}^{n}$. In addition, $\mathbb{S}_+^n$ represents the cone of all symmetric positive semidefinite (PSD) matrices in $\mathbb{S}^n$ with $\mathbb{S}_{++}^n$ denoting its subset of symmetric positive definite (PD) matrices. For any $A, B\in\mathbb{S}_+^n$, the relation $A\succeq B (A \succ B)$ means that $A-B \in\mathbb{S}_+^n (A-B\in\mathbb{S}_{++})$.

\subsection{Problem Setup}
Consider the following discrete-time linear stochastic system:
\begin{equation}\label{sys}
\begin{split}
    x_{t+1} &= A x_t + B u_t + w_t\\
    y_t &= C x_t + v_t,
\end{split}
\end{equation}
where $x_t\in\mathbb{R}^{n_x}$, $u_t\in\mathbb{R}^{n_u}$, and $y_t\in\mathbb{R}^{n_y}$ are the system state, control input, and output at stage $t$, respectively. Here, $w_t\in\mathbb{R}^{n_x}$ represents the system disturbance with unknown distribution, while $v_t\in\mathbb{R}^{n_y}$ is the output noise drawn from a zero-mean Gaussian distribution with covariance matrix $M$. The initial state $x_0$ is also random, drawn from a probability distribution with known mean vector $m_0$ and covariance matrix $M_0$. 
We assume the independence of $w_s$ and $w_t$ and that of $v_s$ and $v_t$ for any $s \neq t$.
Moreover, the random vectors $w_t, v_t$, and $x_t$ are assumed to be independent.

Unlike the fully observable setting, the only information available at time $t$ is the history of noisy measurements $y_0,\dots,y_t$ and the past control inputs $u_0,\dots, u_{t-1}$. Therefore, the information given to the controller at time $t$ can be represented as
\[
\begin{split}
I_t &:=(y_0,\dots, y_t, u_0, \dots, u_{t-1}),\quad t=1,2,\dots,\\
I_0 &:=y_0,
\end{split}
\]
where $I_t$ is called the \emph{information vector}. 
Note that the information vector is updated according to the following dynamical system:
\[
I_{t+1} = (I_t, y_{t+1}, u_t).
\]
In the theory of stochastic optimal control, it is well-known that the information vector serves as a sufficient statistic. 
Thus, it suffices to consider control policies $\pi_t$ that map  $I_t$ to a control input $u_t$ for each $t$.

In many practical problems, the probability distributions of output noise and initial state are given a priori (e.g., known sensor noise). 
In contrast, the distribution of the system disturbances is usually unknown (e.g., unmodelled dynamics).
For simplicity, the disturbance distribution is often assumed to be Gaussian or estimated from data. However, when this assumption is invalid, the imperfect distributional information can deteriorate the controller's performance, especially when it has to operate for an infinite amount of time. Thus, our goal is to design a control policy that is robust against deviations of the true disturbance distribution from the given nominal one. In the literature of DRO, such distributional uncertainties are captured by a set of probability distributions $\mathcal{D}_t \subset \mathcal{P}(\mathbb{R}^{n_x})$, called the~\emph{ambiguity set}. It encompasses prior information about the underlying true distribution and includes distributions with shared structural information. 
As a result, we consider a distribution policy $\gamma_t$ that maps  $I_t$ to a probability distribution $\mathbb{P}_t$ of  $w_t$, chosen from the ambiguity set $\mathcal{D}_t$.

Now, consider the following finite-horizon quadratic cost function:
\begin{equation}\label{unpen_fin_cost}
J_T(\pi,\gamma) := \mathbb{E}_{\bold{y}} \Big[\mathbb{E}_{x_{T}}[x_T^\top Q_f x_T \mid I_{T}] + \sum_{t=0}^{T-1} \mathbb{E}_{x_t} [x_t^\top Q x_t + u_t^\top R u_t \mid I_t, u_t]\Big],
\end{equation} 
where $\pi:=(\pi_0,\dots,\pi_{T-1})$ and $\gamma:=(\gamma_0,\dots, \gamma_{T-1})$, $Q\in\mathbb{S}_+^{n_x}, Q_f\in\mathbb{S}_+^{n_x}, R\in\mathbb{S}_{++}^{n_u}$ are the cost weights, and the outer expectation is taken with respect to the joint distribution of all measurements $\bold{y} := (y_0,\dots, y_{T})$. 
Since our eventual goal is to design a controller  for the infinite-horizon case, we define the following average-cost criterion:
\begin{equation}\label{avg_cost_crit}
J_\infty(\pi,\gamma) = \limsup\limits_{T\to\infty} \frac{1}{T}\mathbb{E}_{\bold{y}} \bigg[\sum_{t=0}^{T-1} \mathbb{E}_{x_t} [x_t^\top Q x_t + u_t^\top R u_t \mid I_t, u_t]\bigg].
\end{equation}

The DRC problem can be formulated as a two-player zero-sum game, where the first player is the controller and the second player is the adversary. The controller selects a policy $\pi=(\pi_0,\pi_1,\dots)$ to minimize the cost, while the adversary player aims to find a distribution policy $\gamma=(\gamma_0,\gamma_1,\dots)$ to maximize the same cost. More precisely, we aim to solve the following minimax stochastic control problem:
\begin{equation}\label{inf_hor}
\min_{\pi\in\Pi}\max_{\gamma\in\Gamma_\mathcal{D}} J_\infty(\pi, \gamma),
\end{equation}
where $\Pi:=\{\pi \mid \pi_t(I_t)=u_t, \pi_t \, \mbox{is measurable} \, \forall t\}$ and $\Gamma_\mathcal{D}:=\{\gamma \mid \gamma_t(I_t)=\mathbb{P}_t \in \mathcal{D}_t, \gamma_t \, \mbox{is measurable} \, \forall t\}$ are the sets of admissible control and distribution policies. Note that the ambiguity set is embedded in the policy space for the adversary, and thus the ambiguity set plays a critical role in characterizing the distributional inaccuracies that are proactively addressed by the controller. 

\subsection{Wasserstein Ambiguity Set}\label{sec:amb_set}

Motivated by the superior properties of Wasserstein DRO mentioned in Section~\ref{sec:intro}, 
we choose $\mathcal{D}_t$ as a Wasserstein ball. 
The Wasserstein metric of order $p$ between two measures $\mathbb{P}$ and $\mathbb{Q}$ supported on $\mathcal{W}\subseteq \mathbb{R}^n$ quantifies the minimum cost of redistributing mass from one measure to another  using non-uniform perturbations and is defined as
\[
W_p(\mathbb{P},\mathbb{Q}):= \inf_{\tau\in \mathcal{T}(\mathbb{P},\mathbb{Q})} \bigg\{\left(\int_{\mathcal{W}^2} \| w - w'\|^p \mathrm{d}\tau(x,y)\right)^{1/p}\bigg\},
\]
where $\mathcal{T}(\mathbb{P} ,\mathbb{Q})$ is the set of all measures in $\mathcal{P}(\mathcal{W}^2)$ with the first and second marginals $\mathbb{P}$ and $\mathbb{Q}$, respectively. Here, $\tau$ is called the \emph{transport plan}, which describes the amount of mass to move from $w$ to $w'$, and $\|\cdot\|$ is a norm on $\mathbb{R}^n$ that measures the transportation cost.

Using the Wasserstein metric of order $p=2$ together with  the standard Euclidean norm, we define the ambiguity set as a ball  of radius $\theta>0$ centered at the given nominal distribution $\mathbb{Q}_t$:
\begin{equation}\label{amb_set}
\mathcal{D}_t:=\{\mathbb{P}_t\in\mathcal{P}(\mathbb{R}^{n_x}) \mid W_2 (\mathbb{P}_t, \mathbb{Q}_t) \leq \theta \}.
\end{equation}
In later sections, 
we show that employing the Wasserstein metric
is useful in partially observable LQ control, as it contributes to obtaining a tractable solution and an out-of-sample performance guarantee, among others. 


\section{Tractable Approximation and Solution}\label{sec:app}

The WDRC problem~\eqref{inf_hor} is difficult to solve for two major reasons. 
First, the Bellman equation for \eqref{inf_hor} involves an infinite-dimensional minimax optimization problem. 
Second, partial observability aggravates the situation because the value (or cost-to-go) function  is defined over the space of the information vectors. 
To resolve these issues, we propose a novel approximation technique and a simple solution to the approximate WDRC problem.
Our method uses a Riccati equation and a tractable SDP problem.

\subsection{Tractable Approximation}\label{sec:tract_approx}

Our approximation technique has two main steps. 
We first introduce an additional penalty term in the cost function, motivated by our previous work for the fully observable case~\cite{kim2022minimax}. 
However, this approximation is insufficient  when the system is partially observable.
Thus, the second step is to further approximate the problem using the Gelbrich bound introduced in~\cite{kuhn2019wasserstein}.

For the first step of the proposed approximation, instead of constraining the adversary player to select a disturbance distribution from the ambiguity set, we penalize the deviation of the distribution $\mathbb{P}_t$ from the nominal distribution $\mathbb{Q}_t$. Specifically, a Wasserstein penalty term is added to the cost function as follows:
\[
\tilde{J}_\infty^\lambda (\pi, \gamma) := \limsup\limits_{T\to\infty} \frac{1}{T}\mathbb{E}_\mathbf{y}\Bigg[\sum_{t=0}^{T-1} \mathbb{E}_{x_t}[x_t^\top  Q x_t  + u_t^\top  R u_t \mid I_t, u_t] - \lambda W_2 (\mathbb{P}_t, \mathbb{Q}_t)^2\Bigg],
\]
where $\lambda >0$ is a user-specified penalty parameter designated for adjusting the conservativeness of the control policy.
Then, the following minimax control problem approximates the original WDRC problem:
\begin{equation}\label{penalty0}
\min_{\pi\in\Pi}\max_{\gamma\in \Gamma} \tilde{J}_\infty^\lambda (\pi, \gamma),
\end{equation}
where the set of admissible distribution policies is defined as $\Gamma:=\{\gamma \mid \gamma_t(I_t) = \mathbb{P}_t\in\mathcal{P}(\mathbb{R}^{n_x})\}$. This set is different from $\Gamma_\mathcal{D}$ in that it does not restrict the distribution $\mathbb{P}_t$ to be selected from the ambiguity set. 
This would give too much freedom to the adversary if there were no penalty terms.
In general, the minimax control problem with the new cost function is intractable due to partial observability and the Wasserstein penalty term. In fully observable settings, when $\mathbb{Q}_t$ is chosen as an empirical distribution, the minimax problem attains a finite-dimensional formulation. However,  problem~\eqref{penalty0} remains intractable  due to partial observability, as demonstrated in Appendix~\ref{app:po}.

The intractability of \eqref{penalty0}  
 motivates the need for another approximation step, where we propose employing the Gelbrich bound introduced in~\cite{kuhn2019wasserstein}.
The Gelbrich bound
is lower than the Wasserstein distance and is valid for any nominal distribution with finite first- and second-order moments.
Let
\begin{equation}\label{means}
\bar{w}_t:=\mathbb{E}_{w_t\sim\mathbb{P}_t}[w_t], \quad \hat{w}_t:=\mathbb{E}_{w_t\sim\mathbb{Q}_t} [w_t ]
\end{equation}
denote the mean vectors of $w_t$ with respect to $\mathbb{P}_t$ and $\mathbb{Q}_t$, respectively. Also, we let
\begin{equation}\label{covs}
\begin{split}
\Sigma_t &:=\mathbb{E}_{w_t\sim\mathbb{P}_t} [(w_t-\bar{w}_t)(w_t - \bar{w}_t)^\top],\\
\hat{\Sigma}_t &:=\mathbb{E}_{w_t\sim\mathbb{Q}_t} [(w_t-\hat{w}_t)(w_t - \hat{w}_t)^\top]
\end{split}
\end{equation}
denote the covariance matrices of $w_t$ with respect to $\mathbb{P}_t$ and $\mathbb{Q}_t$, respectively.
The Gelbrich bound for Wasserstein distance can be described as follows. 

\begin{lemma}\label{lem:gelbrich}
Suppose the mean vectors and covariance matrices
of $\mathbb{P}_t$ and $\mathbb{Q}_t$ are given by~\eqref{means} and~\eqref{covs}, respectively. Then, the following lower-bound holds for the 2-Wasserstein distance:
\begin{equation}\label{bound}
\mathrm{G}(\mathbb{P}_t,\mathbb{Q}_t) := \sqrt{\|\bar{w}_t - \hat{w}_t\|_2^2 + \mathrm{B}^2(\Sigma_t, \hat{\Sigma}_t)}\leq W_2(\mathbb{P}_t,\mathbb{Q}_t),
\end{equation}
where 
\[
\mathrm{B}^2(\Sigma_t, \hat{\Sigma}_t) := \mathrm{Tr}[\Sigma_t + \hat{\Sigma}_t - 2(\hat{\Sigma}_t^{1/2} \Sigma_t \hat{\Sigma}_t^{1/2})^{1/2}].
\]
Furthermore, the inequality holds with equality if $\mathbb{P}_t$ and $\mathbb{Q}_t$ are elliptical with the same density-generating function.
\end{lemma}

The Gelbrich bound relies only on the mean and covariance information, which is a crucial feature for obtaining a tractable solution. 

\begin{remark}
The Gelbrich bound provides a generic lower-bound for the Wasserstein distance for distributions that are not necessarily elliptical. 
Thus, it is applicable to problems with non-Gaussian disturbance distributions. 
The bound discards information about the nominal distribution $\mathbb{Q}_t$ beyond its first- and second-order moments, thereby sacrificing possibly useful information. However, it trades available information for tractability, providing a simple strategy for evaluating the closeness of two distributions. 
In Sections~\ref{sec:perf} and~\ref{sec:stab}, we also show that the resulting controller enjoys various useful theoretical properties despite the limited use of available information.\footnote{The empirical performance of a Gelbrich bound-based approximation has been demonstrated through motion control problems in~\cite{Hakobyan2021map}.} 
\end{remark}

We leverage the Gelbrich bound and define the following  cost function, replacing the Wasserstein penalty term with its lower-bound:
\begin{equation}\label{pen_inf}
J_{\infty}^\lambda(\pi,\gamma) = \limsup\limits_{T\to\infty} \frac{1}{T} \mathbb{E}_\mathbf{y}\Bigg[\sum_{t=0}^{T-1} \mathbb{E}_{x_t}[ x_t^\top  Q x_t + u_t^\top R u_t \mid I_t, u_t] - \lambda \mathrm{G} (\mathbb{P}_t, \mathbb{Q}_t)^2\Bigg].
\end{equation}
Using this cost function, the penalty version~\eqref{penalty0} of the WDRC problem can be approximated as follows:
\begin{equation}\label{infin_approx}
\min_{\pi\in\Pi}\max_{\gamma\in\Gamma} J^\lambda_\infty (\pi,\gamma).
\end{equation}

Having the approximate problem~\eqref{infin_approx}, a closed-form expression of its optimal solution is derived using a Riccati equation
 in the following subsections. We first consider the case of finite-horizon problems and then extend the obtained results to the infinite-horizon average cost setting.

\subsection{Finite-Horizon Problem}\label{sec:fin}

We begin our analysis by first considering the following finite-horizon approximate WDRC problem:
\begin{equation}\label{fin_approx}
    \min_{\pi\in\Pi} \max_{\gamma\in\Gamma} J_T^\lambda(\pi,\gamma),
\end{equation}
where the cost function is defined as
\begin{equation}\label{fin_cost}
J_{T}^\lambda(\pi,\gamma) = \mathbb{E}_\mathbf{y}\bigg[\mathbb{E}_{x_{T}}[x_T^\top Q_f x_T\mid I_{T}] +\sum_{t=0}^{T-1} \Big(\mathbb{E}_{x_t}[x_t^\top Q x_t + u_t^\top R u_t \mid I_t, u_t] - \lambda \mathrm{G} (\mathbb{P}_t, \mathbb{Q}_t)^2\Big)\bigg].
\end{equation}

To solve the minimax problem~\eqref{fin_approx}, we apply the dynamic programming (DP) algorithm by first defining the optimal value function recursively as follows: 
let $V_T(I_T) := \mathbb{E}_{x_T}[x_T^\top Q_f x_T \mid I_T]$ and
\begin{align}
V_t(I_t) := \, & \inf_{u_t\in\mathbb{R}^{n_u}} \sup_{\mathbb{P}_t\in\mathcal{P}(\mathbb{R}^{n_x})} \mathbb{E}_{x_t, y_{t+1}} \Big[x_t^\top Q x_t + u_t^\top R u_t -\lambda \mathrm{G}(\mathbb{P}_t,\mathbb{Q}_t)^2 + V_{t+1} (I_t, y_{t+1}, u_t) \mid I_t, u_t\Big] \nonumber\\
 = \, & \inf_{u_t\in\mathbb{R}^{n_u}} \sup_{\substack{\bar{w}_t \in\mathbb{R}^{n_x},\\ \Sigma_t \in\mathbb{S}_+^{n_x}}} \mathbb{E}_{x_t, y_{t+1}}[x_t^\top Q x_t  + u_t^\top R u_t  \nonumber \\
 &-\lambda [ \| \bar{w}_t - \hat{w}_t \|^2 + \mathrm{B}^2(\Sigma_t, \hat{\Sigma}_t ) ] + V_{t+1} (I_t, y_{t+1}, u_t) \mid I_t, u_t ] \label{bellman}
\end{align}
for $t=T-1, \dots, 0$.
Suppose for a moment that the outer minimization problem has an optimal solution $u_t^*$ and the value function is measurable for every $t$. 
Then, by the DP principle  (e.g.,~\cite{osogami2015robust, saghafian2018ambiguous, Gonzalez2003, Hernandez2012}), we have
\begin{equation}\label{opt_cost}
\inf_{\pi\in\Pi}\sup_{\gamma\in\Gamma} J_T^\lambda (\pi,\gamma) = \mathbb{E}_{y_0}[V_0(I_0)],
\end{equation}
and an optimal control policy $\pi_t^*$ can be constructed using the optimal solutions  of the outer optimization problems for all $t$. 
To this end, we inductively show that the outer minimization problem in the Bellman equation~\eqref{bellman} admits an optimal solution.

Let the expected value of the state $x_t$ conditioned on the information vector $I_t$ under the disturbance distribution generated by the adversary's policy $\gamma$ be denoted by 
\[
\begin{split}
\bar{x}_t &:=\mathbb{E}_{x_t}[x_t \mid I_t].
\end{split}
\]
Also, let
\[
\xi_t := x_t - \bar{x}_t
\]
denote the deviation of the system state from its conditional expectation, and let
\[
\Phi: = B R^{-1} B^\top - \frac{1}{\lambda} I \in \mathbb{S}^{n_x}.
\]

As the first step for our inductive argument, 
we identify an optimal solution to the outer minimization problem in~\eqref{bellman} for time $t$ when $V_{t+1}$ has the following quadratic form. 
\begin{lemma}\label{lem:opt_cont}
Fix $t \in \{0, 1, \ldots, T-1\}$, and
suppose that
\begin{equation}\label{vf_next}
V_{t+1}( I_{t+1})  =  \mathbb{E}_{x_{t+1}}[ x_{t+1}^\top  P_{t+1}x_{t+1}+ \xi_{t+1}^\top S_{t+1} \xi_{t+1} + 2r_{t+1}^\top x_{t+1} \mid I_{t+1}] + q_{t+1},
\end{equation}
for some $P_{t+1}\in\mathbb{S}_{+}^{n_x}, S_{t+1}\in\mathbb{S}_{+}^{n_x}, r_{t+1}\in\mathbb{R}^{n_x}$, and $q_{t+1}\in\mathbb{R}$.
Moreover, assume  that the penalty parameter satisfies $\lambda I \succ P_{t+1}$. Then, the following results hold:
\begin{itemize}
\item The outer minimization problem in~\eqref{bellman} with respect to $u_t$ has the following unique optimal solution:
\begin{equation}\label{u_opt}
u_t^* = K_t \bar{x}_t + L_t,
\end{equation}
where
\begin{align}
K_t &= - R^{-1}B^\top(I+P_{t+1}\Phi)^{-1} P_{t+1} A\label{contr_param_K}\\
L_t &= - R^{-1}B^\top(I+P_{t+1}\Phi)^{-1}  (P_{t+1} \hat{w}_t + r_{t+1})\label{contr_param_L}.
\end{align}

\item 
Given $u_t^*$, the inner maximization problem in~\eqref{bellman} with respect to $w_t$ has the following unique optimal solution:
\begin{equation}\label{mu}
    \bar{w}_t^*  = H_t \bar{x}_t + G_t,
\end{equation}
where
\begin{align}
H_{t} & = (\lambda I - P_{t+1})^{-1} P_{t+1}(A + B K_{t})\label{mean_param_f_1}\\
G_{t} & = (\lambda I - P_{t+1})^{-1} \big(P_{t+1} B L_{t} + r_{t+1} + \lambda\hat{w}_t\big).\label{mean_param_f_2}
\end{align}

\item 
The inner maximization problem in~\eqref{bellman} with respect to $\Sigma_t \in\mathbb{S}_+^{n_x}$ reduces to the following maximization problem:
\begin{equation}\label{max_sigma}
 \max_{\Sigma_t \in\mathbb{S}_+^{n_x} }  \mathbb{E}_{x_{t+1},y_{t+1}}[\xi_{t+1}^\top S_{t+1} \xi_{t+1}\mid I_t] + \mathrm{Tr}[(P_{t+1} - \lambda I) \Sigma_t + 2 \lambda (\hat{\Sigma}_t^{1/2}\Sigma_t\hat{\Sigma}_t^{1/2})^{1/2}].
\end{equation}
\end{itemize}
\end{lemma}

The proof of this lemma can be found in Appendix~\ref{app:opt_cont}.
Using this lemma, we can also show that 
$V_t$ has the same form as $V_{t+1}$
whenever $\lambda I \succ P_{t+1}$.
To preserve the structure of the value function through the Bellman recursion, we impose the following assumption on the penalty parameter, which is also required for the fully observable case~\cite{kim2022minimax}.

\begin{assumption}\label{ass:lambda_ass}
The penalty parameter satisfies $\lambda I \succ P_{t}$ for all $t=1,\dots, T$.
\end{assumption}

Under this assumption, we can use mathematical induction backward in time to recursively show that the value functions $V_t$'s have a specific quadratic form for all $t$ because $V_T = \mathbb{E}_{x_T}[x_T^\top Q_f x_T \mid I_T]$ is already in that form. 
Consequently, it follows from the DP principle that
the optimal control policy can be constructed as follows.

\begin{theorem}\label{thm:sol}
Suppose that Assumption~\ref{ass:lambda_ass} holds and~\eqref{max_sigma} attains an optimal solution. Then, the value function for all $t = 0, \ldots, T$ has the following form:
\[
V_t(I_t) = \mathbb{E}_{x_t}[x_t^\top P_t x_t + \xi_t^\top S_{t}\xi_t + 2 r_t^\top x_t \mid I_t] + q_{t} + \sum_{s=t}^{T-1}z_t(I_t, s).
\]
Here, the coefficients $P_{t}\in\mathbb{S}_{+}^{n_x}, S_{t}\in\mathbb{S}^{n_x}_+, r_{t}\in\mathbb{R}^{n_x}$, and $q_{t}\in\mathbb{R}$ are
found recursively  using the following Riccati equation:
\begin{align}
    P_t  =\, & Q + A^\top (I + P_{t+1}\Phi)^{-1} P_{t+1} A \label{P}\\
    S_t  = \, & Q + A^\top  P_{t+1} A - P_t\label{S}\\
    r_t  =  \, & A^\top (I + P_{t+1}\Phi)^{-1} ( r_{t+1} + P_{t+1}\hat{w}_t)\label{r}\\
    q_t = \, & q_{t+1}  + (2\hat{w}_t - \Phi r_{t+1})^\top(I + P_{t+1}\Phi)^{-1} r_{t+1}+\hat{w}_t^\top (I + P_{t+1}\Phi)^{-1}  P_{t+1} \hat{w}_t - \lambda\mathrm{Tr}[\hat{\Sigma}_t]\label{q} 
\end{align}
with the terminal conditions $P_T = Q_f, S_T = 0, r_T = 0$, and $q_T = 0$.
The term $z_t (I_t, s)$ for $s=t,\dots, T-1$ is given by
\begin{equation}\label{z}
\begin{split}
z_t (I_t, s)  :=  & \sup_{\Sigma_s \in\mathbb{S}_+^{n_x}} \mathbb{E}_{x_{s+1},y_{t+1}, \dots, y_{s+1}}[\xi_{s+1}^\top S_{s+1} \xi_{s+1}\mid I_t] \\
&+ \mathrm{Tr}[(P_{s+1} - \lambda I) \Sigma_s + 2 \lambda (\hat{\Sigma}_s^{1/2}\Sigma_s\hat{\Sigma}_s^{1/2})^{1/2}].
\end{split}
\end{equation}
Moreover, an optimal policy pair can be obtained as follows:
\begin{itemize}
\item
The optimal control policy is uniquely given by
\[
\pi^*_t(I_t) = K_t\bar{x}_t + L_t,
\]
with $K_t$ and $L_t$ defined as~\eqref{contr_param_K} and~\eqref{contr_param_L}, respectively; and

\item For each $I_t$, let $\gamma_t^* (I_t) = \mathbb{P}_t^*$, where $\mathbb{P}_t^*$ is a probability distribution with mean vector defined as~\eqref{mu} and covariance matrix $\Sigma_t^*$ obtained as the maximizer of~\eqref{z} for stage $t$.
Then, $\gamma_t^*$ is an optimal policy for the adversary that generates the worst-case distribution.

\end{itemize}
\end{theorem}

The proof of this theorem can be found in Appendix~\ref{app:sol}.
In the theorem, the existence of $\Sigma_t^*$  is not guaranteed in general. 
However, 
we will see that $\Sigma_t^*$ exists and is obtained in a tractable way 
 if the Kalman filter is used.

It is worth comparing our result with that of the fully observable case~\cite{kim2022minimax}. Due to partial observability, the optimal control policy and the mean vector of the worst-case distribution are affine in the conditional expectation $\bar{x}_t$ instead of the actual state $x_t$. An additional estimator, such as the Kalman filter, is required for computing the state estimates based on the information $I_t$ collected so far. However, the Riccati recursion~\eqref{P}--\eqref{q}, as well as the controller parameters~\eqref{contr_param_K} and~\eqref{contr_param_L}, are independent of the information vector $I_t$. Thus, the \emph{separation principle} holds for our WDRC method, where the state estimation and the optimal control parts can be decoupled, allowing each component to be designed independently.

The standard Kalman filter uses the mean vector and covariance matrix of the ground-truth disturbance distribution.
However, in our problem setting, it is required to estimate the states under disturbances drawn from the worst-case distribution $\mathbb{P}_t^*$. The expected value of $x_{t+1}$ conditioned on $I_t$ is then estimated as follows:
\begin{equation}\label{kalman_mean}
\bar{x}_{t+1} = \bar{x}_{t+1}^{-} + \bar{X}_{t+1}C^\top M^{-1} (y_{t+1} - C \bar{x}_{t+1}^{-}),
\end{equation}
where $\bar{x}_{t+1}^{-} = A \bar{x}_{t} + B u_t^* + \bar{w}_t^*$ with $\bar{x}_{t}^{-} = m_0$. Here, $\bar{X}_{t}$ is the covariance matrix of $x_{t}$ given $I_t$, i.e.,
\[
\bar{X}_t = \mathbb{E}_{x_t}[(x_t - \bar{x}_t)(x_t - \bar{x}_t)^\top \mid I_t],
\]
which can be precomputed by applying the following recursion forward in time:
\begin{align}
    \bar{X}_{t+1} &= \bar{X}_{t+1}^{-} - \bar{X}_{t+1}^{-} C^\top(C \bar{X}_{t+1}^{-} C^\top + M)^{-1}C \bar{X}_{t+1}^{-}\label{kalman_cov}\\
    \bar{X}_{t+1}^{-} &= A \bar{X}_{t} A^\top + \Sigma_t^*,\label{cov_update}
\end{align}
starting from $\bar{X}_{0}^{-} = M_0$.

It follows from Theorem~\ref{thm:sol} and Kalman filter equations~\eqref{kalman_mean}--\eqref{cov_update} that the optimal cost $J_T^\lambda (\pi^*,\gamma^*)$ depends on the worst-case distribution $\mathbb{P}_t^* = \gamma_t^*(I_t)$ only through its first- and second-order moments. Therefore,~\emph{any} distribution with mean vector $\bar{w}_t^*$ and covariance matrix $\Sigma_t^*$ is the worst-case distribution in~\eqref{fin_approx}. If the worst-case distribution is chosen to be Gaussian, then the Kalman filter is an optimal state estimator, as it minimizes the expected mean-squared error of state estimation~\cite{anderson2012optimal}. As stated previously, when the Kalman filter is used for state estimation, the optimization problem~\eqref{z} attains an optimal solution and can be recast as a tractable SDP problem.

\begin{proposition}\label{prop:sdp}
Suppose that the system state at time $t$ is estimated using the Kalman filter given the information vector $I_t$. Then, $z_t(I_t,t)$ given in~\eqref{z} corresponds to the optimal value of the following tractable SDP problem:
\begin{equation}\label{sdp}
\begin{split}
\max_{\substack{X, X^{-} ,\\ Y, \Sigma \in\mathbb{S}_+^{n_x}}} \; & \mathrm{Tr}[S_{t+1} X + (P_{t+1}-\lambda I)\Sigma + 2\lambda Y]\\
\mbox{s.t.} \; &  \begin{bmatrix} \hat{\Sigma}_t^{1/2} \Sigma \hat{\Sigma}_t^{1/2} & Y \\ Y & I\end{bmatrix} \succeq 0\\
&\begin{bmatrix}X^{-} - X & X^{-} C^\top \\ C X^{-} & C X^{-} C^\top + M\end{bmatrix}\succeq 0\\
&C X^{-} C^\top + M \succeq 0\\
& X^{-} = A \bar{X}_t A^\top + \Sigma,
\end{split}
\end{equation}
where $\bar{X}_t$ is the covariance matrix of $x_t$ conditioned on $I_t$.

Moreover, an optimal solution $\Sigma^*$ to the SDP problem~\eqref{sdp}
is the covariance matrix of the worst-case distribution $\mathbb{P}_t^*$ in Theorem~\ref{thm:sol}.
\end{proposition}

The proof of this proposition can be found in Appendix~\ref{app:sdp}. Notably, the reformulated SDP problem~\eqref{sdp} is independent of real-time data such as the measurement $y_t$ and the control input $u_t$. Therefore, the covariance matrix $\Sigma_t^*$ of the worst-case distribution in each time stage can be computed offline by solving the SDP problem~\eqref{sdp} using existing algorithms~\cite{o2016conic, andersen2003implementing, aps2019mosek}. Having the covariance matrix $\Sigma_t^*$, the conditional state covariance matrix $\bar{X}_{t}$ can also be calculated offline by applying the Kalman filter recursion~\eqref{kalman_cov} and~\eqref{cov_update}.
Finally, in order to compute the value function at time $t$, it is sufficient to have $z_s(I_s, s)$ for $s = t, \dots, T-1$ as from the law of total expectation, it follows that $z_t(I_t, s) = z_s(I_s, s), s = t\dots, T-1$.

\subsection{From Finite-Horizon to Infinite-Horizon Problems}\label{sec:inf}

The results obtained for the finite-horizon problem can be extended to the infinite-horizon average cost setting~\eqref{infin_approx} as letting $T$ tend to $\infty$. 
Throughout this subsection, we assume the following:

\begin{assumption}\label{ass:stat}
The nominal distribution $\mathbb{Q}_t$ has a stationary mean vector and a stationary covariance matrix, i.e., $\hat{w}_t\equiv\hat{w}$ and $\hat{\Sigma}_t\equiv \hat{\Sigma}$ for all $t =0, 1, \ldots$.
\end{assumption}

\begin{assumption}\label{ass:stab_obs} $\Phi \succeq 0$, and $(A, \Phi^{1/2})$ is stabilizable and $(A, Q^{1/2})$ is observable.
\end{assumption}

To examine the asymptotic behavior of the recursion~\eqref{P}--\eqref{q}, we first show the convergence of the Riccati equation~\eqref{P} to a steady-state solution $P_{ss}$ of an ARE.

\begin{proposition}\label{prop:P_ss}
Suppose that Assumptions~\ref{ass:lambda_ass}--\ref{ass:stab_obs} hold. Then, there exists a matrix $P_{ss}\in\mathbb{S}_+^{n_x}$ such that for every $P_T \in \mathbb{S}_+^{n_x}$, we have
\begin{equation}\label{conv}
\lim_{T\to \infty} P_t = P_{ss}.
\end{equation}
Furthermore, $P_{ss}$ is the unique symmetric PSD solution of the following ARE:
\begin{equation}\label{are}
P_{ss} = Q + A^\top (I + P_{ss} \Phi)^{-1} P_{ss} A.
\end{equation}
\end{proposition}

The proof of this proposition can be found in Appendix~\ref{app:P_ss}. As a direct consequence, we can show the convergence of $S_t$ and $r_t$ to their corresponding limits.

\begin{lemma}\label{lem:rs_ss}
Suppose that Assumptions~\ref{ass:lambda_ass}--\ref{ass:stab_obs} hold. Then, the matrix $S_t$ and the vector $r_t$ computed recursively according to~\eqref{S} and~\eqref{r} starting from $S_T=0$ and $r_T=0$ converge to
\begin{align}
S_{ss} &= Q + A^\top P_{ss} A - P_{ss},\label{S_ss}\\
r_{ss} &= [I - A^\top (I+P_{ss}\Phi)^{-1}]^{-1} A^\top (I+P_{ss}\Phi)^{-1} P_{ss}\hat{w}\label{r_ss}
\end{align}
as $T\to \infty$, respectively. 
\end{lemma}

The proof of this lemma can be found in Appendix~\ref{app:rs_ss}.
Proposition~\ref{prop:P_ss} and Lemma~\ref{lem:rs_ss} yield to identify the limiting behavior of the finite-horizon optimal policy as the horizon length tends to infinity.

\begin{theorem}\label{thm:steady_state_pol}
Suppose that Assumptions~\ref{ass:lambda_ass}--\ref{ass:stab_obs} hold. Then, as $T\to \infty$, the optimal control policy $\pi_t^* (I_t)$ converges pointwise to the steady-state policy
\begin{equation}\label{steady_state_pol}
\pi_{ss}^*(I_t):=K_{ss}\bar{x}_t + L_{ss},
\end{equation}
where
\begin{align}
K_{ss} &= - R^{-1}B^\top(I+P_{ss}\Phi)^{-1} P_{ss} A,\label{ss_gain}\\
L_{ss} &= - R^{-1}B^\top(I+P_{ss}\Phi)^{-1}  (P_{ss} \hat{w} + r_{ss})\label{ss_bias}.
\end{align}
Furthermore, as $T\to\infty$, the mean vector of the worst-case distribution $\mathbb{P}_t^*$ generated by the adversary converges to
\begin{equation}
   \bar{w}_{t,ss}^*  =  H_{ss} \bar{x}_t + G_{ss},\label{mu_st}
\end{equation}
where
\begin{align}
H_{ss} & = (\lambda I - P_{ss})^{-1} P_{ss}(A + B K_{ss}),\label{mean_param_1}\\
G_{ss} & = (\lambda I - P_{ss})^{-1} \big(P_{ss} B L_{ss} + r_{ss}  + \lambda\hat{w}\big).\label{mean_param_2}
\end{align}
\end{theorem}

The convergence of $K_t, L_t, H_t$, and $G_t$ in~\eqref{contr_param_K}--\eqref{mean_param_f_2} directly follows from the convergence of $P_t$ and $r_t$. The steady-state control policy~\eqref{steady_state_pol} is again affine in the conditional expectation of the system state. However, it is now stationary, making the controller more attractive for practical implementation. 

Theorem~\ref{thm:steady_state_pol} only concerns the mean vector of the worst-case distribution, which is insufficient to analyze the steady-state behavior of the policy $\gamma_t^*$ of the adversary. Therefore, in the remainder of this subsection, we consider a worst-case distribution policy of a special form and show that it is, in fact, optimal to the infinite-horizon average cost problem~\eqref{infin_approx}.
To this end, consider a stationary distribution policy $\gamma_{ss}^*$ that maps the information vector to a probability distribution with the mean vector $\bar{w}_{t,ss}^*$ defined as~\eqref{mu_st} and the stationary covariance matrix $\Sigma_{ss}^*$ defined as an optimal solution to the following maximization problem:
\begin{equation}\label{z_ss_opt}
 \begin{split}
\max_{\substack{X, X^{-},\\ \Sigma \in \mathbb{S}_+^{n_x}}} \;& \mathrm{Tr}[S_{ss} X + (P_{ss} - \lambda I) \Sigma + 2 \lambda (\hat{\Sigma}^{1/2}\Sigma\hat{\Sigma}^{1/2})^{1/2}]\\
 \mbox{s.t.} \; & X^{-} = A X A^\top + \Sigma\\
&X = X^{-} - X^{-} C^\top (C X^{-} C^\top + M)^{-1} C X^{-}.
 \end{split}
 \end{equation}

For further analysis, we impose the following assumption:

\begin{assumption}\label{ass:cont_obs}
$(A,C)$ is detectable and $\left(A, (\Sigma_{ss}^*)^{1/2}\right)$ is stabilizable.
\end{assumption}

It is well known from filtering theory (e.g.,~\cite{kailath2000linear}) that under the distribution policy $\gamma_{ss}^*$ satisfying Assumption~\ref{ass:cont_obs}, the matrix $\bar{X}_{t}^{-}$ given by the recursion in~\eqref{cov_update} tends to a PSD matrix $\bar{X}_{ss}^{-}$ that solves the following filter ARE:
\begin{equation}\label{cov_update_ss}
    \bar{X}_{ss}^{-} = A (\bar{X}_{ss}^{-} - \bar{X}_{ss}^{-} C^\top(C \bar{X}_{ss}^{-} C^\top + M)^{-1}C \bar{X}_{ss}^{-}) A^\top + \Sigma_{ss}^*
\end{equation}
for any initial state covariance matrix $M_0 \in\mathbb{S}^{n_x}_+$.
Consequently, the covariance matrix $\bar{X}_{t}$ converges to the constant PSD matrix 
\begin{equation}\label{ss_cov}
\bar{X}_{ss} = \bar{X}_{ss}^{-} - \bar{X}_{ss}^{-} C^\top (C \bar{X}_{ss}^{-} C^\top + M)^{-1} C \bar{X}_{ss}^{-},
\end{equation}
with the state recursively estimated according to the following asymptotic form:
\begin{equation}\label{ss_update}
\bar{x}_{t+1} = \bar{x}_{t+1}^{-} + \bar{X}_{ss}C^\top M^{-1} (y_{t+1} - C \bar{x}_{t+1}^{-}),
\end{equation}
where $\bar{x}_{t+1}^{-} = A \bar{x}_{t} + B u_t + \bar{w}_{t,ss}^*$ with $\bar{x}_{0|-1} = m_0$.
This property is known as the duality between estimation and control. As a result, the asymptotic performance of the filter is similar to that of the standard Riccati equation, yielding the steady-state counterpart of the Kalman filter.

Due to its constraints, the optimization problem~\eqref{z_ss_opt} is intractable. Using a similar argument to Proposition~\ref{prop:sdp}, \eqref{z_ss_opt} can be reformulated as the following tractable SDP problem:
 \begin{equation}\label{sdp_ss}
 \begin{split}
\max_{X, X^{-}, Y, \Sigma \in \mathbb{S}_+^{n_x}} \; & \mathrm{Tr}[S_{ss} X + (P_{ss}-\lambda I)\Sigma_{ss} + 2\lambda Y]\\
 \mbox{s.t.} \; &  \begin{bmatrix} \hat{\Sigma}^{1/2} \Sigma \hat{\Sigma}^{1/2} & Y \\ Y & I\end{bmatrix} \succeq 0\\
 &\begin{bmatrix}X^{-} - X & X^{-} C^\top \\ C X^{-} & C X^{-} C^\top + M\end{bmatrix}\succeq 0\\
 &C X^{-} C^\top + M \succeq 0\\
 & X^{-} = A X A^\top + \Sigma,
 \end{split}
 \end{equation}
which is independent of the information vector $I_t$ and can be solved offline.

Finally, we can build the connection between the policy pair $(\pi_{ss}^*, \gamma_{ss}^*)$ and the solution to the infinite-horizon minimax problem~\eqref{infin_approx}. For that, let the steady-state average cost incurred by the stationary policy pair $(\pi_{ss}^*, \gamma_{ss}^*)$ be denoted as
 \[
 \rho:= J_\infty^\lambda (\pi_{ss}^*, \gamma_{ss}^*),
 \]
 which can be calculated by combining the results from Theorem~\ref{thm:sol} and the maximization problem~\eqref{z_ss_opt} as follows.

\begin{proposition}\label{prop:avg_cost}
Suppose that Assumptions~\ref{ass:lambda_ass}--\ref{ass:cont_obs} hold. Then, the steady-state average cost is given by
\begin{equation}\label{avg_cost}
\rho  =  (2\hat{w} - \Phi r_{ss})^\top ( I+P_{ss}\Phi)^{-1}r_{ss}- \lambda\mathrm{Tr}[\hat{\Sigma}]+ \hat{w}^\top (I+P_{ss}\Phi)^{-1}P_{ss}\hat{w}  + z_{ss},
\end{equation}
where $z_{ss}$ is the optimal value of the maximization problem~\eqref{z_ss_opt}.
\end{proposition}

The proof of this proposition can be found in Appendix~\ref{app:avg_cost}. 
Having the steady-state average cost, it remains to verify the optimality of the policy pair $(\pi_{ss}^*, \gamma_{ss}^*)$ in the average-cost criterion. For that purpose, we introduce the following optimality condition:

 \begin{proposition}\label{prop:bellman}
Suppose that Assumptions~\ref{ass:lambda_ass}--\ref{ass:cont_obs}
 hold. Then, the following average-cost optimality equation holds:
 \begin{equation}\label{bellman_1}
 \rho + h(I_t) =  \inf_{u_t\in\mathbb{R}^{n_u}} \sup_{\mathbb{P}_t\in\mathcal{P}(\mathbb{R}^{n_x})} \mathbb{E}_{x_t, y_{t+1}} \Big[x_t^\top Q x_t + u_t^\top R u_t-\lambda \mathrm{G}(\mathbb{P}_t,\mathbb{Q}_t)^2 + h (I_{t+1}) \mid I_t, u_t\Big],
 \end{equation}
 where $\rho$ is the steady-state average cost defined as~\eqref{avg_cost} and
 \[
 h(I_t) = \bar{x}_t^\top P_{ss} \bar{x}_t + 2 r_{ss}^\top \bar{x}_t + \mathrm{Tr}[(S_{ss} + P_{ss})\bar{X}_{ss}].
 \]
 In addition, $(\pi_{ss}^*(I_t), \gamma_{ss}^*(I_t))$ is an optimal solution pair to the minimax problem on the right-hand side of~\eqref{bellman_1}.
 \end{proposition}

The proof of this proposition can be found in Appendix~\ref{app:bellman}. Here, $h$ is called the \emph{bias} and represents the transient cost.
 Using the bias term, we now consider the following extended average-cost function: 
\begin{equation}\label{ext_cost}
\bar{J}^\lambda_\infty (\pi, \gamma):= \limsup_{T\to\infty} \frac{1}{T}\bar{J}_T^\lambda(\pi,\gamma),
\end{equation}
where
\begin{equation}\label{ext}
\bar{J}_T^\lambda (\pi,\gamma) = \mathbb{E}_\mathbf{y}\left[h(I_T) +  \sum_{t=0}^{T-1} \mathbb{E}_{x_t}[x_t^\top Q x_t + u_t^\top R u_t \mid I_t] - \lambda \mathrm{G} (\mathbb{P}_t, \mathbb{Q}_t)^2  \right].
\end{equation}
The extended average cost~\eqref{ext_cost} allows us to investigate the optimality of the steady-state policy pair $(\pi_{ss}^*,\gamma_{ss}^*)$.

\begin{proposition}\label{prop:minmax_opt}
Suppose that Assumptions~\ref{ass:lambda_ass}--\ref{ass:cont_obs} hold. Then, the steady-state policy pair $(\pi_{ss}^*,\gamma_{ss}^*)$ is optimal to
    \[
    \min_{\pi\in\bar{\Pi}}\max_{\gamma\in\bar{\Gamma}} J^\lambda_{\infty}(\pi,\gamma)
    \]
    for any policy spaces $\bar{\Pi}\subset \Pi$ and $\bar{\Gamma}\subset \Gamma$ satisfying
    \begin{align}
        &\limsup\limits_{T\to\infty} \frac{1}{T} \mathbb{E}_\mathbf{y}[ h(I_T) \mid \pi, \gamma_{ss}^*] = 0, \; \forall \pi\in\bar{\Pi} \label{cond_1}\\
        &\limsup\limits_{T\to\infty} \frac{1}{T} \mathbb{E}_\mathbf{y}[ h(I_T) \mid \pi_{ss}^*, \gamma] = 0, \;  \forall \gamma\in\bar{\Gamma}.\label{cond_2}
    \end{align}
    Moreover, the optimal value of this problem is equal to $\rho$.
\end{proposition}

The proof of this proposition can be found in Appendix~\ref{app:minmax_opt}. 
The first condition is similar to the one in the standard LQG control, with the difference that the disturbances follow the worst-case distribution policy $\gamma_{ss}^*$. If the expected value of the state with respect to all uncertainties is bounded under the policy pair $(\pi_{ss}^*,\gamma)$ for some $\gamma\in\bar{\Gamma}$, then condition~\eqref{cond_2} holds. In fact, condition~\eqref{cond_2} is satisfied as long as the distribution $\mathbb{P}_t = \gamma (I_t)$ has a bounded mean vector and a stationary covariance matrix so that the pair $(A,\Sigma^{1/2})$ is stabilizable. This is due to the stability properties of the optimal control policy $\pi_{ss}^*$, which is discussed in Section~\ref{sec:stab}.

We wrap up this subsection observing the tightness of the proposed Gelbrich bound-based approximation when the nominal distribution $\mathbb{Q}_t$ is elliptical.
This is because the worst-case distribution can be chosen to be elliptical with the worst-case mean vector and covariance matrix.

\begin{proposition}\label{prop:ellipt}
Suppose that the nominal distribution $\mathbb{Q}_t$ is elliptical for all $t$. Let $(\pi^*, \gamma^*)$ denote an optimal policy pair of the approximate minimax control problem~\eqref{infin_approx}, such that the worst-case distribution $\mathbb{P}_t^*=\gamma_t^*(I_t)$ is elliptical with the same density generating function as $\mathbb{Q}_t$. Then, $(\pi^*, \gamma^*)$ is an optimal policy pair for the minimax control  problem~\eqref{penalty0}.
\end{proposition}
The proof of this proposition can be found in Appendix~\ref{app:ellipt}. This property once again confirms the validity of our approximation scheme, as most LQ optimal control problems use  nominal distributions as Gaussian. 
For general distributions, 
 the proposed approximate controller is further shown to have performance guarantees in Section~\ref{sec:perf}.

\subsection{Algorithm}

\begin{algorithm}[t]
\DontPrintSemicolon
\SetKw{Input}{Input:}
\SetKw{to}{to}
\SetKw{and}{and}
\SetKw{break}{break}
\Input $\lambda, \hat{w}, \hat{\Sigma}, m_0, M$\;

Solve ARE~\eqref{are} to obtain $P_{ss}$\;
Calculate $K_{ss}$ and $L_{ss}$ by~\eqref{ss_gain} and~\eqref{ss_bias}\;
Compute parameters $H_{ss}$ and $G_{ss}$ according to~\eqref{mean_param_1} and~\eqref{mean_param_2}\;
Solve SDP problem~\eqref{sdp_ss} to obtain $\Sigma_{ss}^*$\;
Solve filter ARE~\eqref{cov_update_ss} and use~\eqref{ss_cov} to obtain $\bar{X}_{ss}$\;
Measure $y_0$ and estimate $\bar{x}_0$ via~\eqref{ss_update}\;
\For{$t=0,1,\dots$}{
Apply $u_t^* = \pi_{ss}^*(I_t) = K_{ss}\bar{x}_t + L_{ss}$ to the system~\eqref{sys}\;
Compute the worst-case mean $\bar{w}_{t,ss}^*$ according to~\eqref{mu_st}\;
Measure $y_{t+1}$ and estimate $\bar{x}_{t+1}$ via~\eqref{ss_update}\;
}
\caption{Infinite-horizon WDRC algorithm} \label{alg:PO_WDRC}
\end{algorithm}

The results presented in previous sections lead us to a novel infinite-horizon WDRC  scheme that controls the partially observable system~\eqref{sys} while continuously updating the state estimates. The block diagram of our method is depicted in Fig.~\ref{fig:overview}, while the detailed procedure is given in Algorithm~\ref{alg:PO_WDRC}. 
The penalty parameter $\lambda$ is initially given to the algorithm, chosen depending on the desired level of conservativeness and satisfying Assumption~\ref{ass:lambda_ass}. 
The remaining inputs of the algorithm include the mean vector $\hat{w}$ and the covariance matrix $\hat{\Sigma}$ of the nominal distribution $\mathbb{Q}_t$, the initial state mean vector $m_0$, and the covariance matrix of the output noise $M$. Our algorithm essentially comprises two stages: offline and online, where the first stage concerns the controller and estimator design, while the second stage is for real-time deployment of the controller.

Since the separation principle applies to our method, we disentangle the controller from the state estimator. Therefore, in the first part, a stationary optimal control policy is synthesized (Lines 2 and 3), followed by the worst-case distribution policy construction (Lines 4 and 5). More specifically, in Line 2, the ARE~\eqref{are} is solved to obtain the matrix $P_{ss}$, which is used in Line 3 to calculate  $K_{ss}$ and $L_{ss}$ according to~\eqref{ss_gain} and~\eqref{ss_bias}, respectively. Next, in Line 4, the parameters $H_{ss}$ and $G_{ss}$ of the mean vector of the worst-case disturbance distribution are found according to~\eqref{mean_param_1} and~\eqref{mean_param_2}, respectively. 
In Line 5, the SDP problem~\eqref{sdp_ss} is solved numerically using the steady-state matrices $P_{ss}$ and $S_{ss}$. Next, in Line 6, we solve the filter ARE~\eqref{cov_update_ss} and~\eqref{ss_cov} to obtain the conditional state covariance matrix $\bar{X}_{ss}$ under the worst-case distribution.

The online stage for the fixed controller and estimator is presented in Lines 7--11, where the optimal policy $\pi_{ss}^*$ is deployed to control the actual partially observable system. In the beginning, an initial measurement $y_0$ is received, and the initial state estimate $\bar{x}_0$ is obtained by the Kalman filter (Line 7). Then, in each time stage, a control input $u_t^*$ is applied to the system leveraging the optimal policy $\pi_{ss}^*$ and the current state estimate $\bar{x}_t$ (Line 9). The mean vector $w_{t,ss}^*$ of the worst-case distribution is then computed according to~\eqref{mu_st} using the parameters $H_{ss}$ and $G_{ss}$ calculated in the offline stage. Finally, in Line 11, the new measurements $y_{t+1}$ are used to update the estimate about the state $x_{t+1}$.

\section{Performance Guarantees}\label{sec:perf}

Though our approach yields a closed-form expression for the optimal control policy of the approximate minimax control problem~\eqref{infin_approx}, its relation to the original WDRC problem~\eqref{inf_hor} is yet to be established. In this section, we demonstrate the capability of our method to provide distributional robustness with a guaranteed cost property and a probabilistic out-of-sample performance guarantee, which is an essential feature of the WDRC method.

\subsection{Guaranteed Cost Property}

Fix a penalty parameter $\lambda > 0$ satisfying Assumption~\ref{ass:lambda_ass}. The corresponding solution to ARE~\eqref{are} will be $P_{ss}$. Now, consider the average cost criterion~\eqref{avg_cost_crit} and its extended version with the bias $h$ being added as follows:
\begin{equation}\label{unpen_avg_cost_ext}
\bar{J}_\infty (\pi,\gamma) = \limsup\limits_{T\to\infty} \frac{1}{T}\mathbb{E}_\mathbf{y}\bigg[h(I_T) +  \sum_{t=0}^{T-1}  \mathbb{E}_{x_t}[x_t^\top Q  x_t  +  u_t^\top R u_t \mid I_t, u_t] \bigg].
\end{equation}
The following theorem demonstrates the uniform bound on the average-cost criterion~\eqref{avg_cost_crit} under the stationary control policy computed in Theorem~\ref{thm:sol} for any worst-case distribution in the Wasserstein ambiguity set $\mathcal{D}$. 

\begin{theorem}\label{thm:guarantee}
Suppose that Assumptions~\ref{ass:lambda_ass}--\ref{ass:cont_obs} hold for a fixed $\lambda > 0$. Also, let $\pi_{ss}^{\lambda,*}$ be the optimal control policy of the penalty problem~\eqref{infin_approx}. 
For any policy space $\bar{\Gamma}_\mathcal{D}$ defined in Proposition~\ref{prop:minmax_opt}, 
the average cost under the worst-case distribution policy in  $\bar{\Gamma}_\mathcal{D}$ is bounded as follows:
\begin{equation}\label{avg_cost_ub}
\sup_{\gamma\in\bar{\Gamma}_\mathcal{D}} J_\infty (\pi_{ss}^{\lambda,*},\gamma) \leq  \theta^2 \lambda+ \rho(\lambda).
\end{equation}
\end{theorem}

The proof of this theorem can be found in Appendix~\ref{app:guarantee}. Theorem~\ref{thm:guarantee} demonstrates the distributional robustness of the optimal control policy $\pi_{ss}^{\lambda,*}$ to the approximate penalty problem, which can be controlled by tuning $\lambda$. The bound~\eqref{avg_cost_ub} suggests an intuitive approach for selecting the penalty parameter given a Wasserstein ball radius $\theta$, as it is desirable to select a $\lambda$ that minimizes the upper-bound,\footnote{This approach was used to determine $\lambda$ for our experiments in Section~\ref{sec:exp}.}  i.e.,
\begin{equation}\label{eq:lambda}
\lambda (\theta)  \in \argmin_{\lambda > 0} \; [ \theta^2\lambda + \rho (\lambda)].
\end{equation}
This optimal penalty parameter is used in the following subsection.

\subsection{Out-of-Sample Performance Guarantee}

Suppose that the standard stochastic optimal controller is constructed using an empirical disturbance distribution constructed from 
the training dataset $\bold{w} := \{\hat{w}^{(1)},\dots,\hat{w}^{(N)}\}$.
The performance of this controller is deteriorated when evaluated under a testing dataset of $w_t$ which is different from the training dataset. 
This issue arises even if the training and testing datasets are sampled from the same disturbance distribution. 
A substantial advantage of WDRC is to address this out-of-sample issue by providing a performance guarantee~\cite{yang2020wasserstein}.

We argue that such an out-of-sample performance guarantee is achieved by the proposed method despite approximation. 
Specifically, we show that for a well-calibrated Wasserstein ambiguity set,
our method with a nominal  empirical distribution provides an upper confidence bound on the true average cost.
Throughout this section, the nominal distribution is chosen as the following
stationary empirical  distribution $\mathbb{Q}$  constructed from a finite sample dataset $\bold{w}$:
\begin{equation}\label{eq:emp}
\mathbb{Q}=\frac{1}{N}\sum_{i=1}^{N} \delta_{\hat{w}^{(i)}},
\end{equation}
where $\delta_w$ denotes the Dirac measure concentrated at $w$. Here, each sample $\hat{w}^{(i)}$ is drawn from the true stationary distribution $\mathbb{P}$.

Given the optimal penalty parameter $\lambda (\theta)$ defined as~\eqref{eq:lambda},
let $(\pi_{ss, \bold{w}}^{\lambda (\theta), *}, \gamma_{ss, \bold{w}}^{\lambda (\theta), *})$ denote the optimal stationary policy pair constructed in Section~\ref{sec:inf} with the sample dataset $\mathbf{w}$. Then, the out-of-sample performance (or cost) of $\pi_{ss, \bold{w}}^{\lambda (\theta), *}$ is defined as
\[
J_\infty(\pi_{ss, \bold{w}}^{\lambda (\theta), *},\gamma)= \limsup\limits_{T\to\infty}\frac{1}{T} \mathbb{E}_\mathbf{y}\bigg[\sum_{t=0}^{T-1}\mathbb{E}_{x_t}[x_t^\top Q x_t + u_t^\top R u_t \mid I_t, u_t] \: \bigg \vert \:\pi_{ss, \bold{w}}^{\lambda (\theta), *}, \gamma\bigg],
\]
where $\gamma$ is a stationary policy mapping the information vector to the true disturbance distribution, i.e., $\gamma(I_t) = \mathbb{P}$  for all $t$.

However, as the true distribution $\mathbb{P}$ is unknown in practice, it is impossible to directly evaluate the out-of-sample performance. Instead, we consider the following alternative probabilistic performance guarantee:
\begin{equation}\label{perf_guar}
\mathbb{P}^{N}\left\{\bold{w} \mid J_\infty(\pi_{ss, \bold{w}}^{\lambda (\theta), *},\gamma) \leq \theta^2 \lambda(\theta)  +\rho(\lambda(\theta)) \right\} \geq 1-\beta,
\end{equation}
where $\beta\in(0,1)$ represents a confidence level. Here, the dataset $\bold{w}$ is viewed as a random object governed by the distribution $\mathbb{P}^{N}$.
The inequality~\eqref{perf_guar} means that
the cost incurred by the proposed policy under the true disturbance distribution
is limited by $\theta^2 \lambda(\theta)+ \rho (\lambda(\theta))$ with probability no less than $1-\beta$. 
Note that the cost upper-bound $\theta^2\lambda(\theta) + \rho (\lambda(\theta))$  can be computed using the proposed method without the knowledge of the true distribution $\mathbb{P}$. 
The probability on the left-hand side critically depends on $\theta$.
Thus, given $\beta$, the size of the ambiguity set must be carefully determined to attain the probabilistic out-of-sample performance guarantee. 

We identify the desired radius $\theta$ under 
the following assumption, ensuring that $\mathbb{P}$ is a light-tailed distribution:
\begin{assumption}\label{ass:light}
Suppose there exist $c>2$ and $B>0$ such that
\[
\mathbb{E}_{w\sim \mathbb{P}}[\exp (\|w\|^c)]\leq B.
\]
\end{assumption}

The required radius $\theta$ can then be found from the following measure concentration inequality for the Wasserstein metric~\cite[Theorem 2]{Fournier2015}:
\begin{equation} \label{measure}
\mathbb{P}^{N} \big\{ \mathbf{w}  \mid 
W_2  (\mathbb{P},\mathbb{Q}) \geq \theta
\big\}\leq c_1 \big [ b_1(N, \theta) \bold{1}_{\{\theta\leq 1\}} + b_2(N, \theta) \bold{1}_{\{\theta > 1\}} \big ],
\end{equation}
where
\begin{equation} \nonumber
b_1 (N, \theta) := \left \{
\begin{array}{ll}
\exp (-c_2 N \theta^2) & \mbox{if } n_x < 4\\
\exp \big(-c_2 N \big(\frac{\theta}{\log(2+1/\theta)}\big)^2 \big) & \mbox{if } n_x = 4\\
\exp (-c_2 N \theta^{n_x/2}  ) &\mbox{otherwise}
\end{array}
\right.
\end{equation}
and
\[
b_2 (N, \theta) := 
\exp ( -c_2 N \theta^{c/2} )
\]
for some constants $c_1, c_2 > 0$, depending only on $n_x$ and $c$.
The measure concentration inequality~\eqref{measure} provides an upper-bound on the probability that the true disturbance distribution $\mathbb{P}$ lies outside the Wasserstein ambiguity set $\mathcal{D}$. This inequality is essential for determining the radius $\theta$ required for ensuring the probabilistic out-of-sample performance of our control policy.

\begin{theorem}\label{thm:perf_guar}
Suppose that Assumptions~\ref{ass:lambda_ass}--\ref{ass:light} hold. We also assume that the radius $\theta$ is chosen as
\begin{equation}\label{radius}
\theta := \begin{cases} \Big[ \frac{\log (c_1 / \beta)}{c_2 N} \Big]^{2/c} & \text{if}\; N < \frac{1}{c_2}\log (c_1/\beta)\\ \Big[ \frac{\log(c_1 / \beta)}{c_2 N} \Big]^{1/2} & \text{if}\; N \geq\frac{1}{c_2} \log(c_1/\beta),\;n_x<4\\ \Big[ \frac{\log(c_1 / \beta)}{c_2 N} \Big]^{2/n_x} & \text{if}\; N \geq\frac{1}{c_2} \log(c_1/\beta),\;n_x>4\\ \bar{\theta} & \text{if}\; N \geq\frac{(\log 3)^2}{c_2} \log(c_1/\beta),\;n_x=4\end{cases}
\end{equation}
for $\bar{\theta}$ satisfying the condition
 \[
 \frac{\bar{\theta}}{\log (2+1/\bar{\theta})}=\bigg[\frac{\log (c_1/\beta)}{c_2 N}\bigg]^{1/2}.
 \]
Then, the probabilistic out-of-sample performance guarantee~\eqref{perf_guar} holds.
\end{theorem}

The proof of this theorem can be found in Appendix~\ref{app:perf_guar}.

Under an additional assumption that the disturbance distribution $\mathbb{P}$ is compactly supported, the concentration inequality suggested in~\cite[Proposition 3.2]{Boskos2020} can be used to further strengthen our result. 
Let the diameter of a set $S \in \mathbb{R}^{n_x}$ be denoted by $\mathrm{diam}(S): = \sup\{\|x-y\|_\infty \mid x, y \in S\}$, and for $\mathbb{P}\in\mathcal{P}(\mathbb{R}^{n_x})$ let $\mathrm{supp}(\mathbb{P})$ denote its support.

\begin{cor}
Suppose that Assumptions~\ref{ass:lambda_ass}--\ref{ass:cont_obs} hold and the true disturbance distribution $\mathbb{P}$ is compactly supported with $\xi:=\frac{1}{2} \mathrm{diam}(\mathrm{supp}(\mathbb{P}))$. Suppose the radius $\theta$ is chosen as
\[
\theta := \begin{cases} \xi \Big[ \frac{\log(c_1 / \beta)}{c_2 N} \Big]^{1/4} & \text{if}\; n_x<4\\ \xi \Big[ \frac{\log(c_1 / \beta)}{c_2 N} \Big]^{1/n_x} & \text{if}\; n_x>4\\ \bar{\theta} & \text{if}\; n_x=4\end{cases}
\]
for $\bar{\theta}$ satisfying the condition
 \[
 \frac{\bar{\theta}^2}{\xi^2 \log (2+\xi^2/\bar{\theta}^2)}=\bigg[\frac{\log (c_1/\beta)}{c_2 N}\bigg]^{1/2},
 \]
 where $c_1, c_2 >0$ are some constants depending only on $n_x$. Then, the probabilistic out-of-sample performance guarantee~\eqref{perf_guar} holds.
\end{cor}

\section{Stability}\label{sec:stab}

This section investigates the stability properties of the closed-loop system when the proposed control policy $\pi_{ss}^*$ is employed. 
It follows from Theorem~\ref{thm:steady_state_pol} that 
the closed-loop system is expressed as
\begin{equation}\label{closed_loop_sys} 
x_{t+1} = Ax_t + BK_{ss} \bar{x}_t + w_t + B L_{ss},
\end{equation}
where $\bar{x}_t$ is the current state estimate. 
Assuming that the Kalman filter is chosen as the state estimator, 
our focus is to analyze the following mean-state system:
\begin{equation}\label{mean_sys}
\begin{split}
\mathbb{E}[x_{t+1}] &= A\mathbb{E}[x_t] + BK_{ss}\mathbb{E}[\bar{x}_t] + \mathbb{E}[w_t] + B L_{ss}\\
\mathbb{E}[\bar{x}_{t+1}]  &=  \mathbb{E}[\bar{x}_{t+1}^-] + \bar{X}_{ss} C^\top M^{-1} C  \mathbb{E}[x_{t+1} - \bar{x}_{t+1}^-]\\
\mathbb{E}[y_t] &= C\mathbb{E}[x_t] + \mathbb{E}[v_t],
\end{split}
\end{equation}
where $\mathbb{E}[\bar{x}_{t+1}^-] =  (A + B K_{ss} + H_{ss})\mathbb{E}[\bar{x}_t] + BL_{ss} + G_{ss}$.
Here, the expectation is taken with respect to the joint probability distribution of all uncertainties up to time $t$.

Let
\[
\tilde{x}_t := \mathbb{E}[x_t], \quad \bbar{x}_t:= \mathbb{E}[\bar{x}_t]
\]
consist of the state of the mean-state system~\eqref{mean_sys}.
We can show the stabilizing properties of the policy pair $(\pi_{ss}^*,\gamma_{ss}^*)$ for the mean-state system when the nominal disturbance distribution $\mathbb{Q}_t$ has zero mean.
\begin{proposition}\label{prop:stability}
Suppose that Assumptions~\ref{ass:lambda_ass}--\ref{ass:cont_obs} hold. Under the  policy pair $(\pi_{ss}^*, \gamma_{ss}^*)$, both $\tilde{x}_t$ and  $\bbar{x}_t$ of the mean-state system~\eqref{mean_sys} converge to the following value: 
\begin{equation}\label{conv_state}
[I - (I+\Phi P_{ss})^{-1} A]^{-1} (I - \Phi(I+P_{ss}\Phi - A^\top)^{-1} P_{ss}] \hat{w}.
\end{equation}
Moreover, if $\hat{w} = \mathbb{E}_{w_t \sim \mathbb{Q}_t} [ w_t ] =0$, the control policy $\pi_{ss}^*$ stabilizes the system under the worst-case distribution policy $\gamma_{ss}^*$.
\end{proposition}

The proof of this proposition can be found in Appendix~\ref{app:stability}. Furthermore, we can show that $\pi_{ss}^*$ guarantees the BIBO stability of the closed-loop system~\eqref{mean_sys} when viewing the disturbances as input.

\begin{proposition}\label{prop:bibo}
Suppose that Assumptions~\ref{ass:lambda_ass}--\ref{ass:stab_obs} hold and the pair $(A,C)$ is detectable. Then, the closed-loop gain matrix $(A+B K_{ss})$ is stable. Moreover, the mean-state system~\eqref{mean_sys} under the control policy $\pi_{ss}^*$ is BIBO stable when viewing the disturbances as input. 
\end{proposition}

The proof of this proposition can be found in Appendix~\ref{app:bibo}. It follows from BIBO stability that as long as the mean vector of the disturbance distribution is bounded, the expected value of the closed-loop system state  and the corresponding output will remain bounded.

\section{Case Study}\label{sec:exp}

In this section, the performance of our  WDRC method is evaluated on a power system frequency regularization problem using the IEEE 39 bus system, which models the New England power grid~\cite{kim2022minimax}. The linearized second-order model for power  systems has the following form:
\begin{equation}\label{sys_power}
\begin{bmatrix}
\Delta \dot{\delta} \\ \Delta \dot{\omega}
\end{bmatrix} = \begin{bmatrix}
0 & I \\ - \bar{M}^{-1} \bar{L} & - \bar{M}^{-1} \bar{D}
\end{bmatrix}\begin{bmatrix}\Delta \delta \\ \Delta \omega\end{bmatrix}
+ \begin{bmatrix} 0 \\ \bar{M}^{-1} \end{bmatrix} \Delta P,
\end{equation}
where $\bar{M}$ and $\bar{D}$ are the diagonal matrices of inertia and
damping coefficients, $\bar{L}$ is the Laplacian matrix of the transmission network. The system state $x(t) := [\Delta \delta^\top (t), \Delta \omega^\top(t)]^\top\in\mathbb{R}^{20}$ consists of the rotor angles and frequencies for 10 generators, while the control input $u(t):=\Delta P(t)\in\mathbb{R}^{10}$ is the power injection vector of the generators. 
It is assumed that only the rotor angle and frequency of the first six generators are measured, i.e., $n_y = 12$ with $C = [I_{12 \times 6}, \mathbf{0}_{12 \times 4}, I_{12 \times 6}, \mathbf{0}_{12 \times 4}]$.
The continuous-time system~\eqref{sys_power} is discretized by a zero-order hold method with sample time $0.1$ seconds. 
This yields a discrete-time stochastic system model of the form~\eqref{sys}.
A disturbance $w(t)$ drawn from an unknown distribution affects the power system dynamics. Such disturbances arise from fluctuations in  net demand, mechanical noise in generators, etc. We test our method with two different disturbance distributions: $(i)$ Gaussian and $(ii)$ uniform. The performance of our method is compared with that of the standard infinite-horizon LQG controller~\cite{bertsekas2012dynamic}. Since the true disturbance distribution is unknown, LQG directly uses the nominal distribution in both the controller and the estimator. The results are obtained by running the algorithms for $100$ time steps. All the experiments were implemented in Python and run on a PC with Intel Core i7-8700K @ 3.70 GHz CPU and 32 GB RAM. The source code of our infinite-horizon WDRC implementation, as well as its finite-horizon version, is available online.\footnote{\href{https://github.com/CORE-SNU/PO-WDRC}{\tt https://github.com/CORE-SNU/PO-WDRC}}

\subsection{Gaussian Case}

\begin{figure}[t]
     \centering
     \begin{subfigure}[b]{0.4\linewidth}
         \centering
         \includegraphics[width=\linewidth]{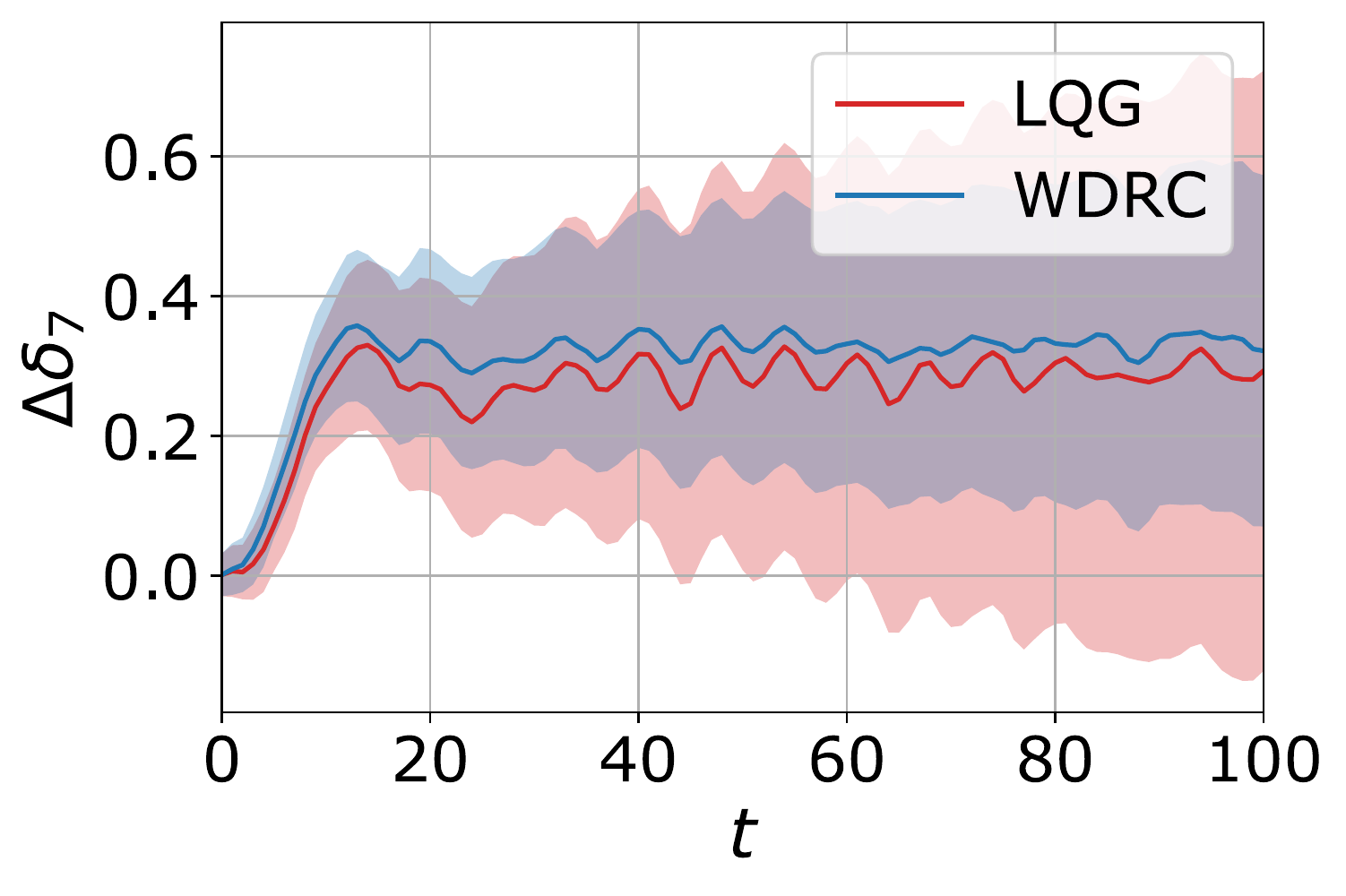}
         \caption{ }
         \label{fig:state_power_1}
     \end{subfigure}%
     \begin{subfigure}[b]{0.4\linewidth}
         \centering
         \includegraphics[width=\linewidth]{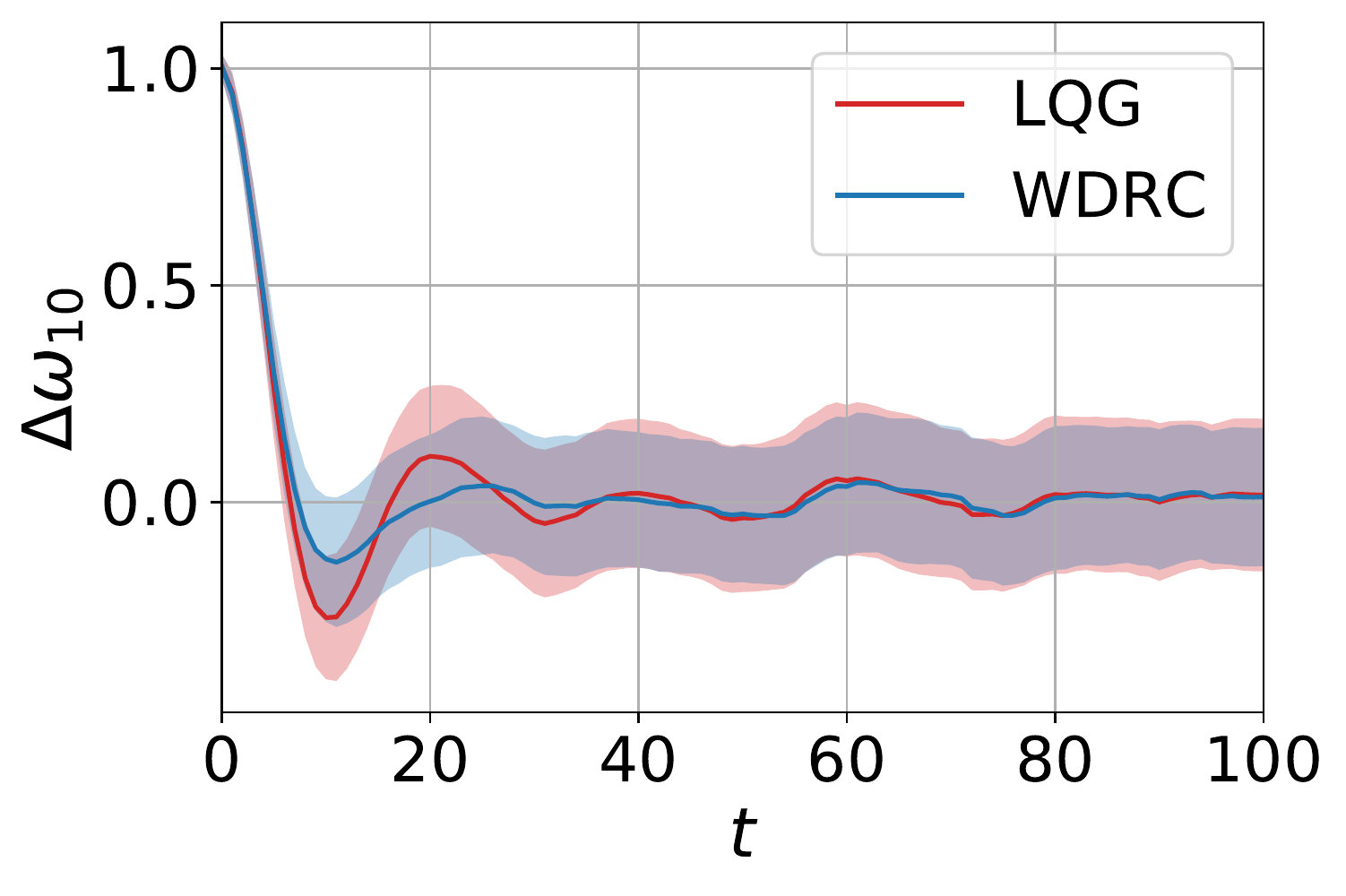}
         \caption{ }
         \label{fig:state_power_2}
	 \end{subfigure}
    \caption{Trajectories of $\Delta \delta_7$ and $\Delta \omega_{10}$ for the system controlled by the LQG and WDRC methods averaged over 1,000 simulation runs in the case of Gaussian disturbances. The shaded regions represent $25\%$ of the standard deviation.}
    \label{fig:state_n}
\end{figure}

\begin{figure*}[t]
     \centering
     \begin{subfigure}[b]{0.45\linewidth}
         \centering
         \includegraphics[width=\linewidth]{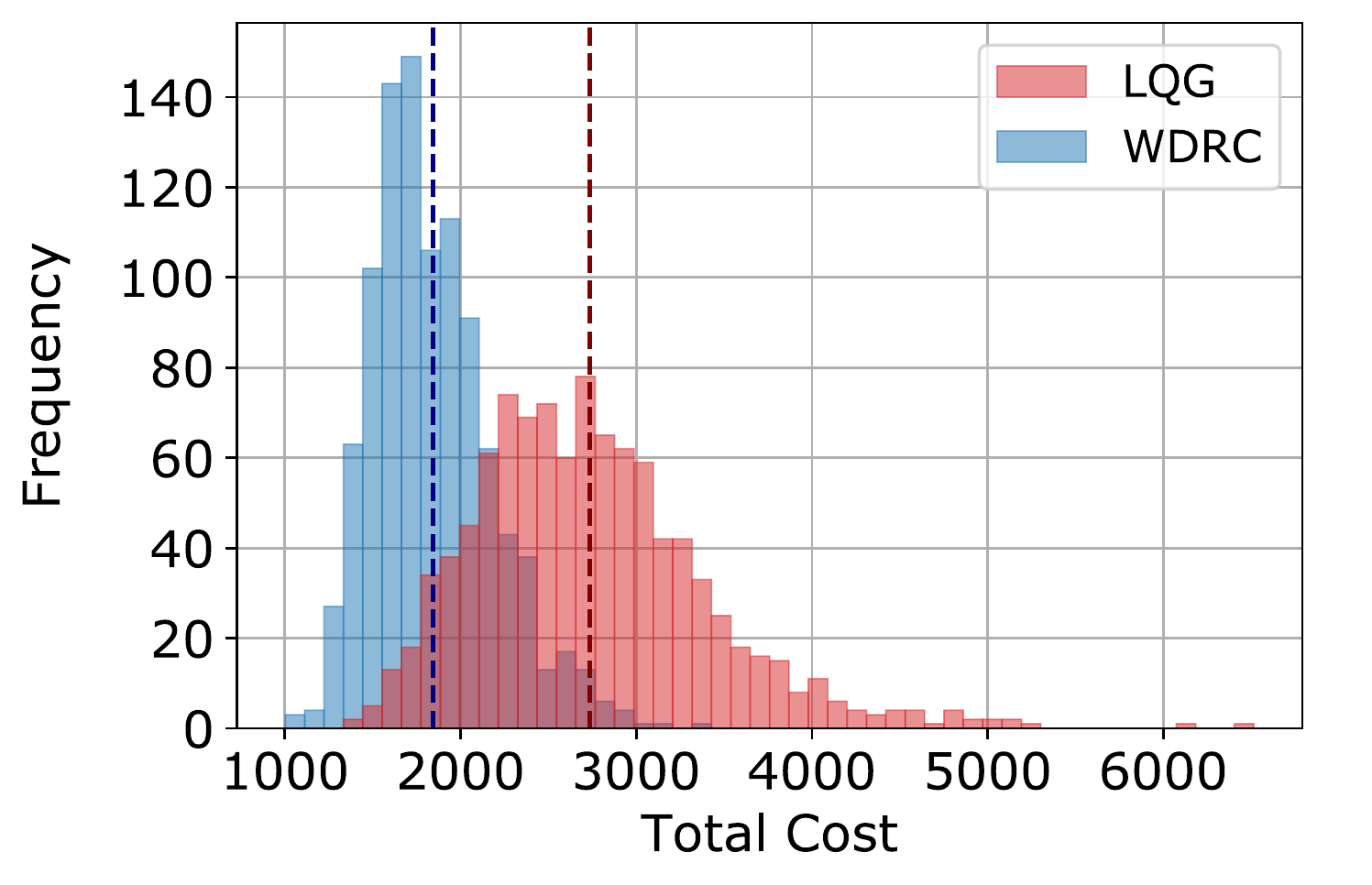}
         \caption{ }
         \label{fig:hist_n}
     \end{subfigure}%
     \begin{subfigure}[b]{0.45\linewidth}
         \centering
         \includegraphics[width=\linewidth]{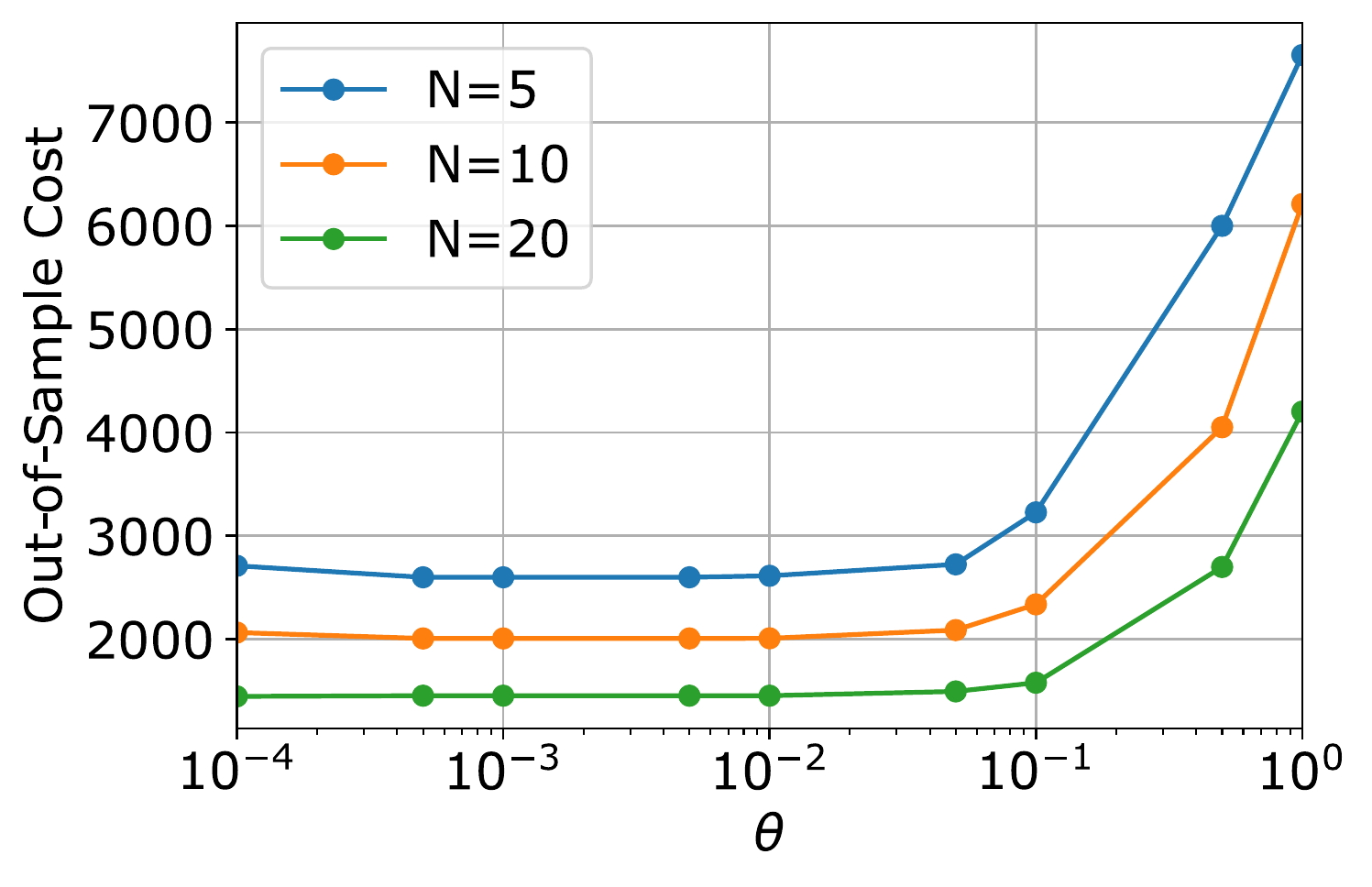}
         \caption{ }
         \label{fig:os_n}
	 \end{subfigure}
    \caption{(a) Histogram of the total costs incurred by the LQG and WDRC methods; and (b) out-of-sample performance of WDRC in the case of Gaussian disturbances. }
    \label{fig:normal}
\end{figure*}

In these experiments, the initial state distribution is Gaussian with mean $m_0 = [\mathbf{0}_{19}, 1]^\top$ and covariance matrix $M_0 = 0.01 I_{20}$. The true disturbances are drawn from a zero-mean Gaussian distribution with a covariance matrix $\Sigma = 0.01 I_{20}$, while the observation noise has a covariance $M=0.01 I_{12}$. The nominal distribution is constructed using $N=5$ disturbance sample data by letting $\hat{\mu}=0$ and $\hat{\Sigma}$ be the empirical covariance matrix. We select the penalty parameter $\lambda$ by minimizing the upper-bound in~\eqref{avg_cost_ub} for $\theta=10^{-3}$.

Fig.~\ref{fig:state_n} displays the state trajectories of $\Delta \delta_{7}$ and $\Delta \omega_{10}$, which are both unobservable states, controlled by the WDRC and LQG methods. The results are averaged over 1,000 simulation runs. These results indicate that the WDRC method reduces the fluctuations and the large variance in the rotor angle and removes unnecessary undershoot in the frequency. Besides, our method successfully keeps the states stable despite the inaccurate nominal distribution.
The total cost and the computation time for running the whole algorithm are reported in Table~\ref{table:cost_comp}. The WDRC method yields a lower average total cost with a smaller variance over the simulations than the LQG method. Furthermore, the computation times for running the two methods are almost identical, as the computationally expensive SDP problem and the Riccati equations are solved in the offline stage, making the complexity of the online stage similar for both algorithms.

\begin{table}[t]
\caption{Total cost and online computation time averaged over 1,000 simulations.}
\centering
\setlength{\tabcolsep}{0.2em} 
\begin{tabular}{>{\raggedright}m{2.2cm} >{\centering}m{3.2cm}  >{\centering\arraybackslash}m{3.2cm}  >{\centering}m{3.2cm}  >{\centering\arraybackslash}m{3.2cm}}
\toprule
 & \multicolumn{2}{c}{\textbf{Total Cost}} & \multicolumn{2}{c}{\textbf{Computation Time}}\\
\cmidrule(lr){2-3}  \cmidrule(lr){4-5}
& WDRC & LQG & WDRC & LQG\\
\midrule
\textbf{Gaussian} & 1842.640 (341.836) & 2735.015 (661.369) & 0.113 (0.019) & 0.115 (0.014) \\
\cline{2-5}
\textbf{Uniform} & 1891.211 (394.855) & 2653.224 (767.445) & 0.0184 (0.003) & 0.0183 (0.002)\\
\bottomrule
\end{tabular}
\label{table:cost_comp}
\end{table}

Fig.~\ref{fig:normal} (a) displays the distribution of total costs computed for 1,000 simulation runs. It reveals that for WDRC,  the overall distribution is concentrated in the low-cost region. In contrast, the total costs induced by the LQG controller are comparatively higher as it relies on the nominal disturbance distribution, disregarding possible inaccuracies due to the small sample size. Meanwhile, our WDRC method penalizes deviations of the true distribution from the nominal one, thereby making the controller more robust against distributional uncertainties.

Fig.~\ref{fig:normal} (b) shows the out-of-sample cost incurred by our method for different values of the ambiguity set radius $\theta$ and various sample sizes of the dataset $\mathbf{w}$ estimated for 10,000 disturbance samples drawn from the true distribution and averaged over 1,000 independent simulation runs. For each $\theta$, the penalty parameter $\lambda(\theta)$ is found according to~\eqref{eq:lambda}. We observe that  the cost slightly decreases as the radius increases up to $\theta=10^{-3}$. The cost starts growing for $\theta \in \{10^{-3}, 10^{0}\}$. This is because a large $\theta$ encourages the controller to be overly conservative, while the controller with a small $\theta$ is not sufficiently robust.  

\begin{figure}[t]
    	\centering	\includegraphics[width=0.45\linewidth]{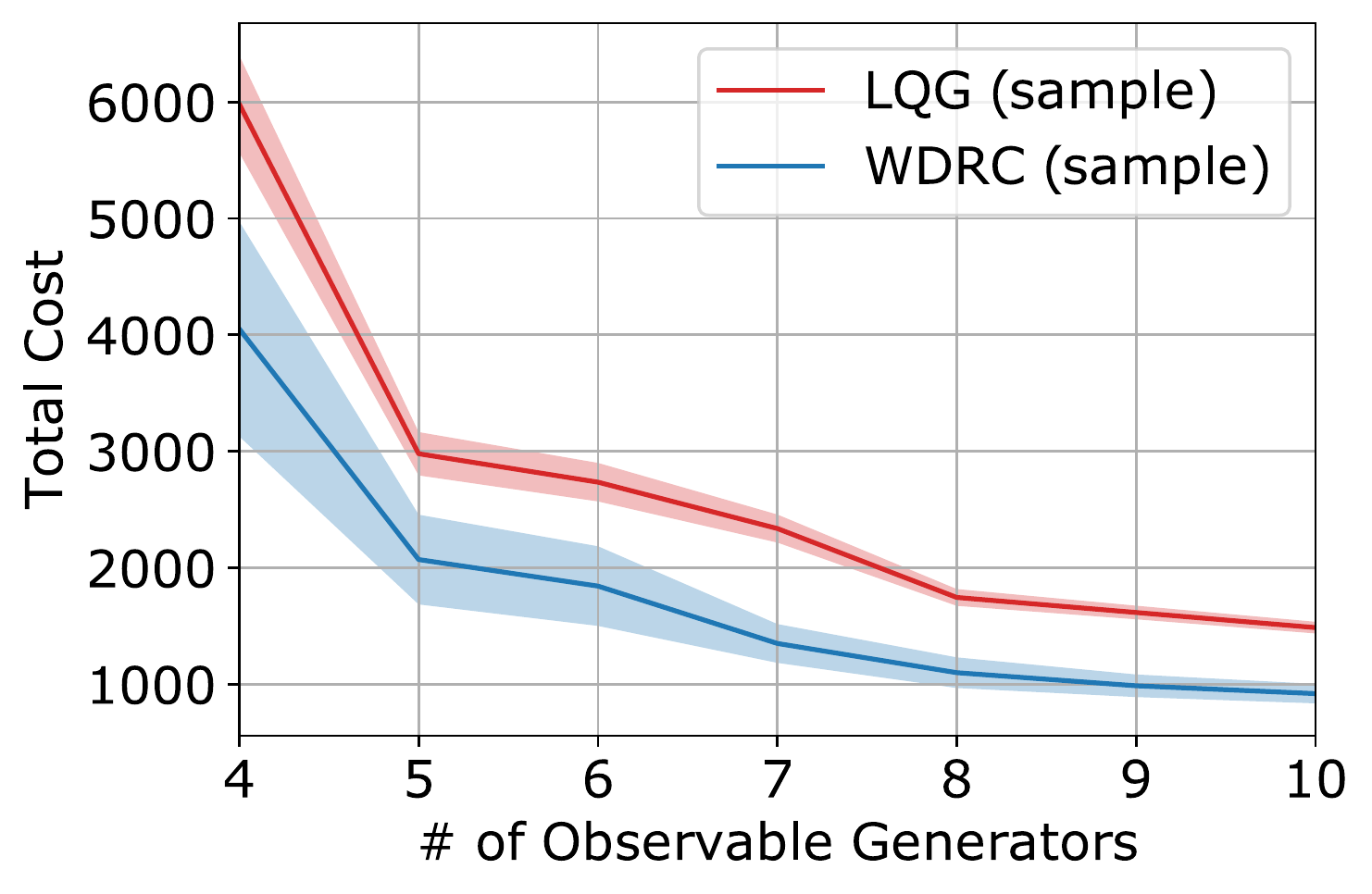}
    \caption{Effect of the number of  observable generators on the total cost incurred by the LQG and WDRC methods averaged over 1,000 simulation runs in the case of normal disturbances. The shaded regions represent $25\%$ of the standard deviation.}
    \label{fig:cost_obs}
\end{figure}

As part of these experiments, we also examine the effect of partial observability on the control performance. Specifically, Fig.~\ref{fig:cost_obs} shows the total costs incurred by the WDRC and LQG methods under a varying number of observable generators. It can be seen that regardless of the number of observable generators, our method outperforms  LQG. Overall, the total cost decreases as more generators become observable, resulting in smaller mean and variance values. 

\subsection{Uniform Case}

In this scenario, the true disturbances in each dimension follow a uniform distribution $\mathcal{U}(-0.15, 0.15)$. The initial state distribution is also uniform, $\mathcal{U}(-0.05,0.05)$ for all states, except $\Delta\omega_{10}$, for which the initial state is selected from $\mathcal{U}(0.95,1.05)$. The nominal distribution is constructed using $N=5$ sample data drawn from the true distribution with its mean and covariance corresponding to the empirical ones. The penalty parameter is chosen by minimizing the upper-bound in~\eqref{avg_cost_ub} for $\theta=10^{-2}$.

The Kalman filter is an optimal estimator only in the Gaussian case. However, we approximate the disturbance distribution by a Gaussian, assuming that $w_t^*\sim \mathcal{N}(\bar{w}_{t,ss}^*, \Sigma_{ss}^*)$, and apply the steady-state Kalman filter. Besides, unlike the usual LQG settings, where the observation noise is assumed to be zero-mean Gaussian, we draw it from a uniform distribution $\mathcal{U}(-0.4, 0.4)$ and estimate the covariance matrix from $40$ samples. By doing so, we  evaluate the capability of our WDRC algorithm in the presence of an erroneous state estimator.

\begin{figure}[t]
     \centering
     \begin{subfigure}[b]{0.4\linewidth}
         \centering
         \includegraphics[width=\linewidth]{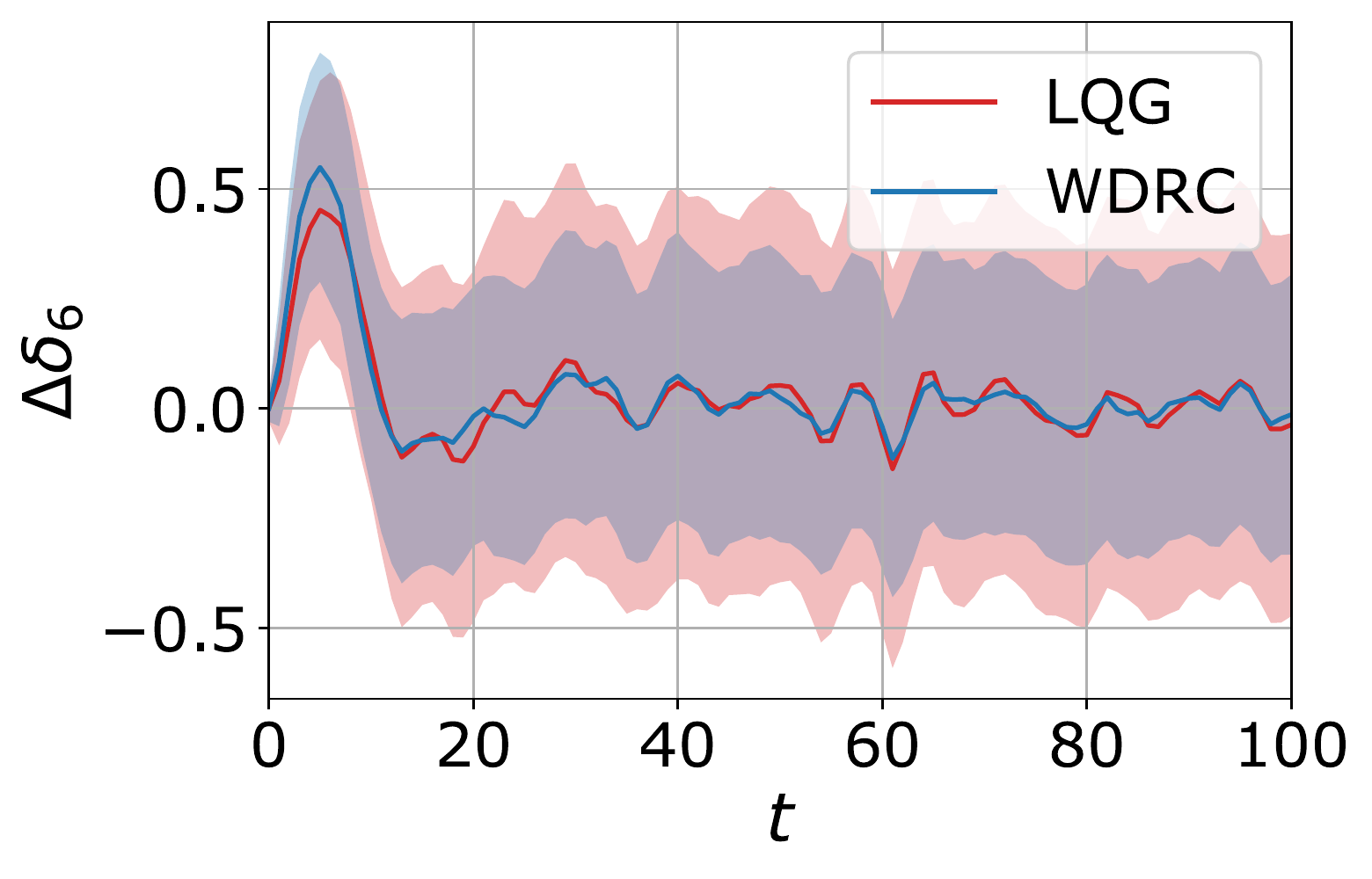}
         \caption{ }
    \label{fig:hist_u}
     \end{subfigure}%
     \begin{subfigure}[b]{0.4\linewidth}
         \centering
         \includegraphics[width=\linewidth]{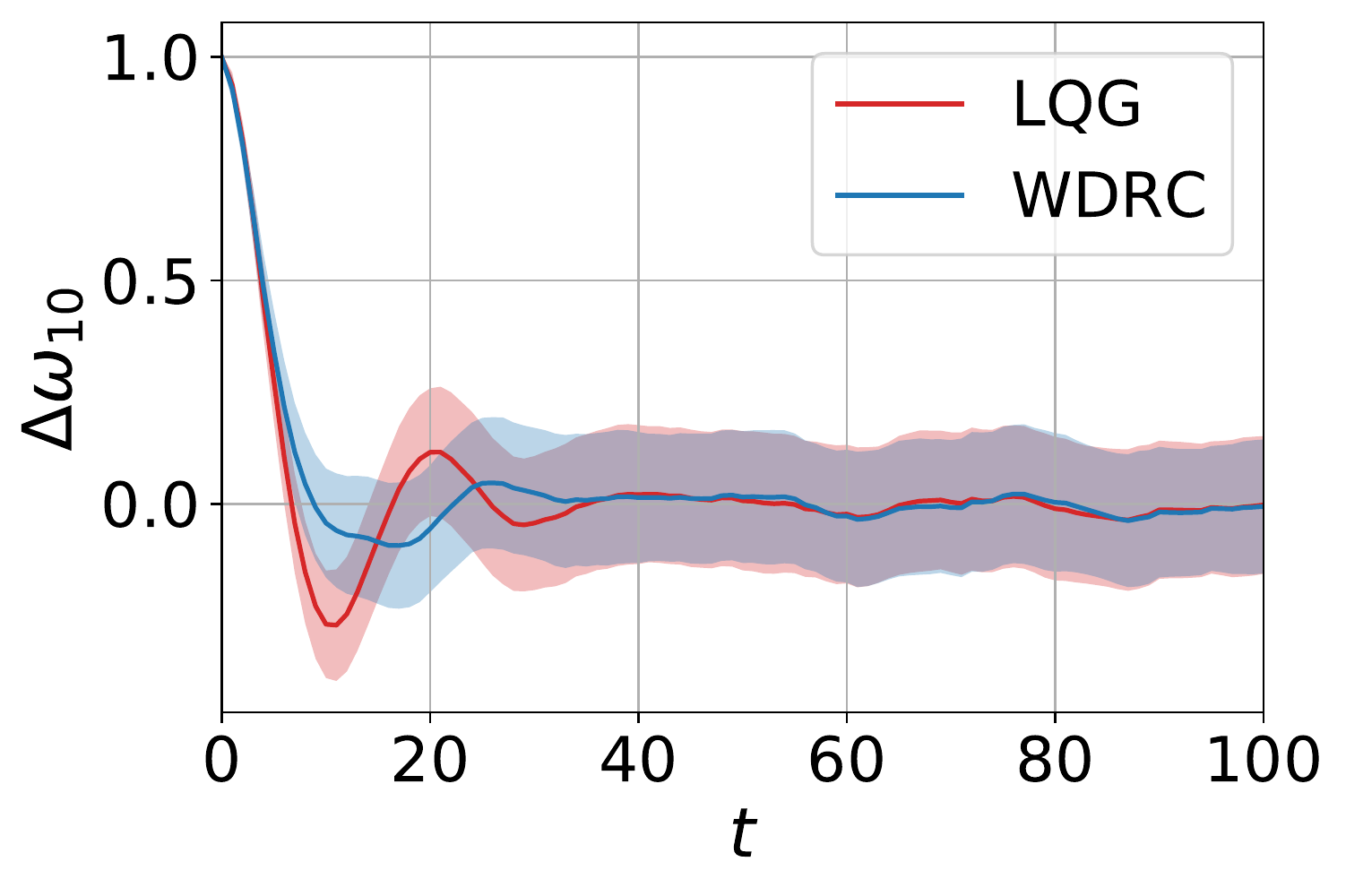}
         \caption{ }
         \label{fig:state_m_2}
     \end{subfigure}
    \caption{Trajectories of $\Delta \delta_6$ and $\Delta \omega_{10}$ for the system controlled by the LQG and WDRC methods averaged over 1,000 simulation runs in the case of uniform disturbances. The shaded regions represent $25\%$ of the standard deviation.}
    \label{fig:state_u}
\end{figure}

Fig.~\ref{fig:state_u} displays the state trajectories for $\Delta \delta_6$ and $\Delta \omega_{10}$ for the WDRC and LQG methods averaged over 1,000 simulation runs. 
It shows that LQG results in a larger variance in the trajectory for $\Delta \delta_6$, which is reduced in the WDRC case. In addition, our method smooths the unwanted fluctuations in the trajectory of $\Delta\omega_{10}$ present in the LQG case. The total cost and the computation time for running the algorithm are presented in Table~\ref{table:cost_comp}. Our WDRC method outperforms the LQG method in total cost, inducing a lower average cost with a smaller variance.

\begin{figure*}[t]
     \centering
     \begin{subfigure}[b]{0.45\linewidth}
         \centering
         \includegraphics[width=\linewidth]{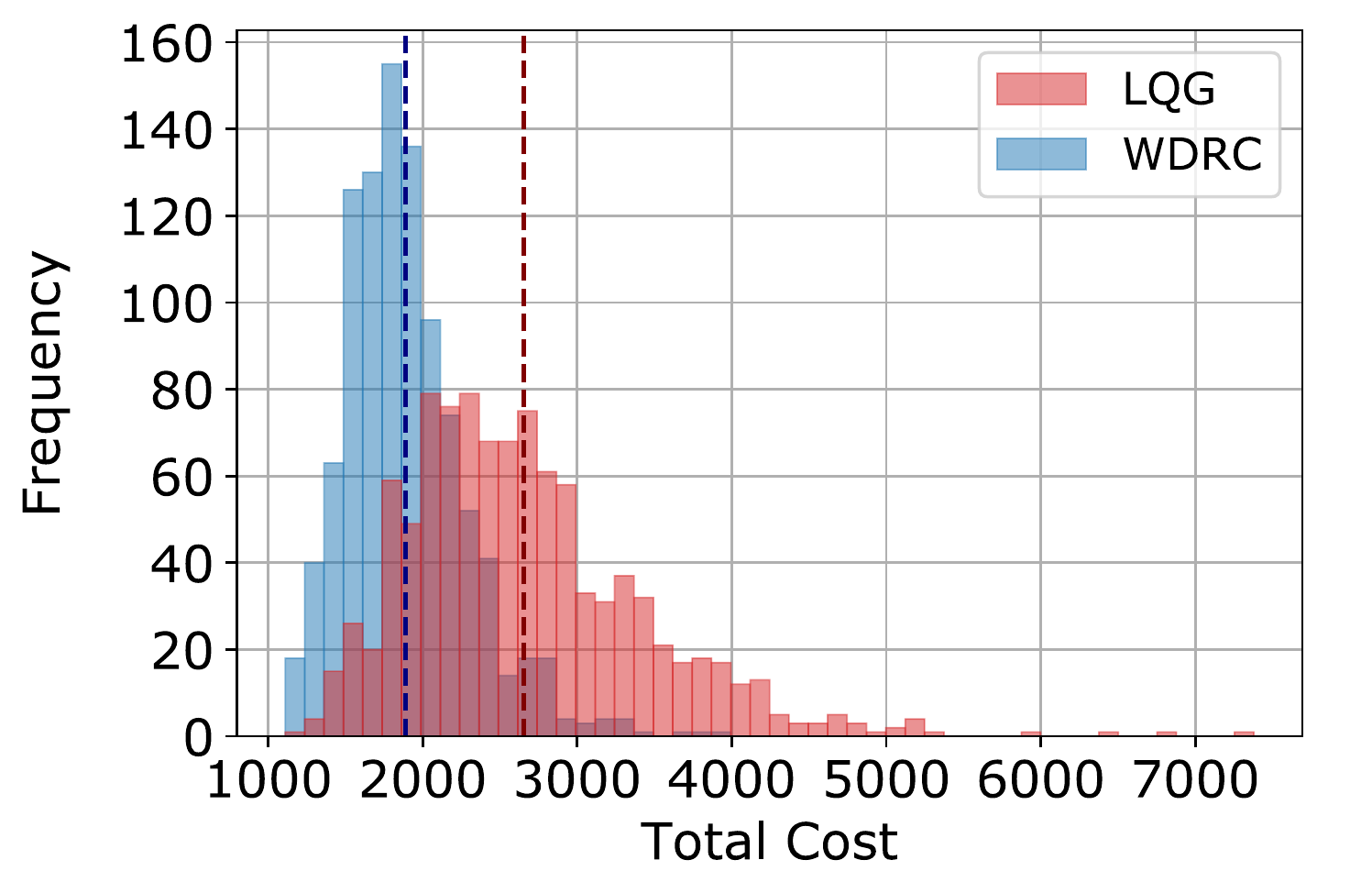}
         \caption{ }
         \label{fig:hist_u}
     \end{subfigure}%
     \begin{subfigure}[b]{0.45\linewidth}
         \centering
         \includegraphics[width=\linewidth]{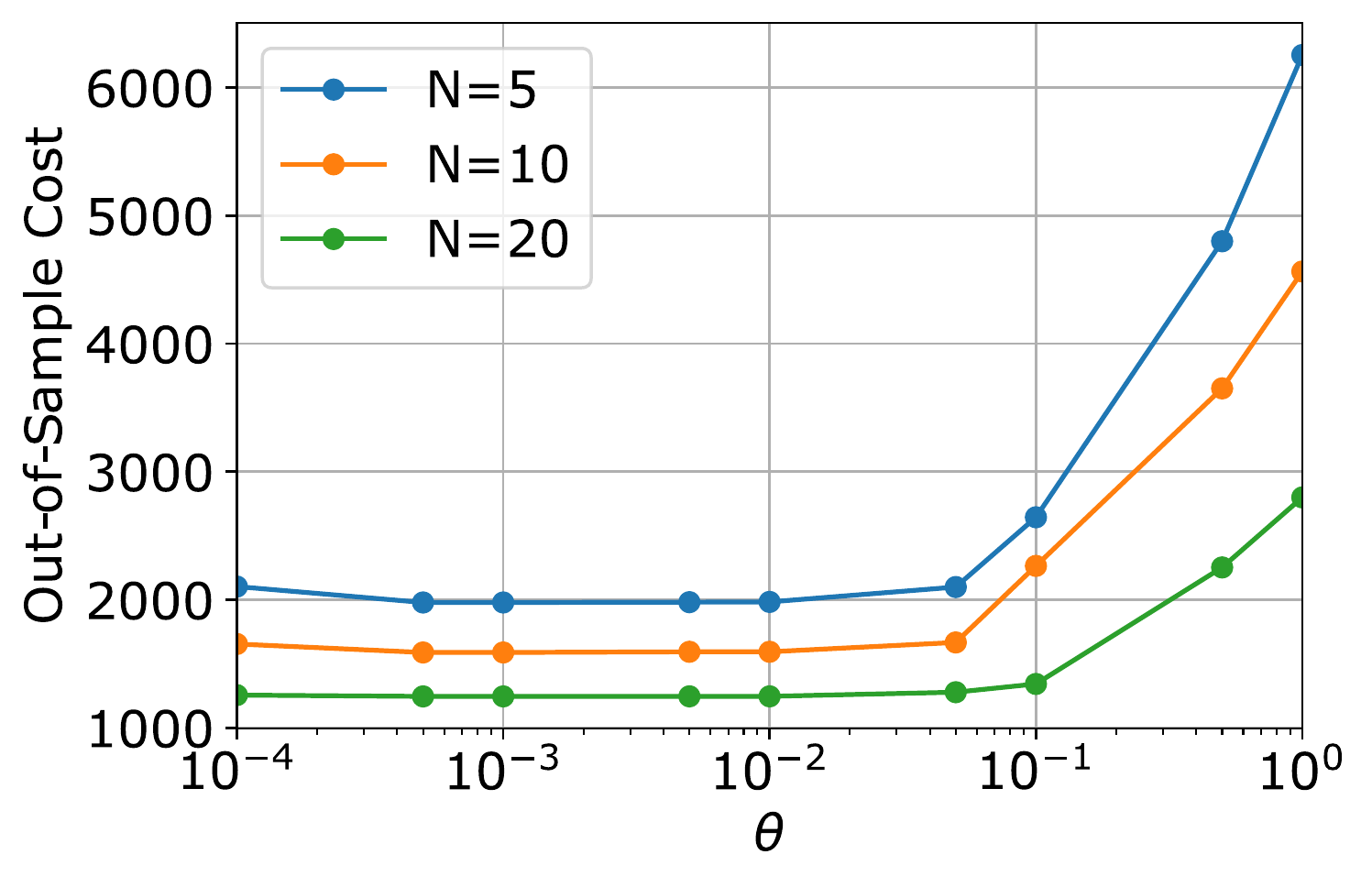}
         \caption{ }
         \label{fig:os_u}
	 \end{subfigure}
    \caption{(a) Histogram of the total costs incurred by the LQG and WDRC methods, and (b) out-of-sample cost of WDRC in the case of uniform disturbances.}
    \label{fig:uniform}
\end{figure*}

The distribution of total costs computed for 1,000 simulation runs is presented as a histogram in Fig.~\ref{fig:uniform} (a). Overall, the total costs incurred by WDRC are smaller than the ones induced by the LQG method. Furthermore, the costs for applying the proposed method are concentrated in the low-cost region, whereas the cost distribution for LQR is relatively widespread, covering a large range of costs. This happens because the LQG controller is designed solely using the mean and covariance of the nominal distribution. Furthermore, the state estimation is performed for an inaccurate disturbance distribution, aggravating the situation. Our WDRC method resolves these issues by considering the worst-case disturbance distribution close to the nominal one, thereby anticipating mismatches between the actual and nominal distributions during both the control and estimation stages.

Fig.~\ref{fig:uniform} (b) illustrates the total out-of-sample cost induced by our method for different values of $\theta$ and $N$ estimated for 10,000 disturbance samples drawn from the true distribution. The results are averaged over 1,000 independent simulation runs. Similar to the previous scenario, the cost slightly decreases as the radius increases up to $\theta=10^{-3}$ and the cost increases thereafter. 

\begin{figure}[t]
    	\centering
    	\includegraphics[width=0.45\linewidth]{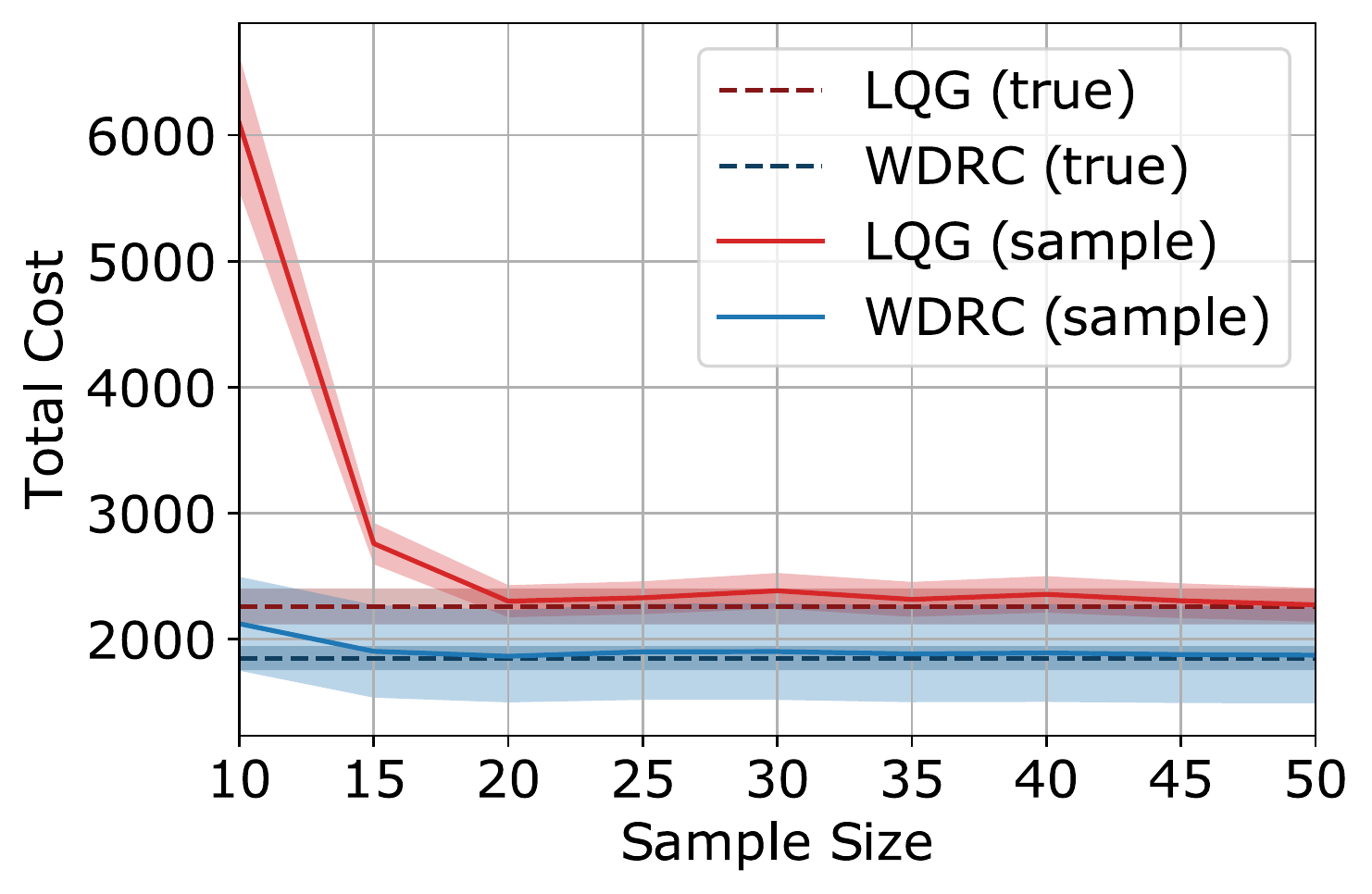}
    \caption{Effect of measurement noise uncertainty on the total cost incurred by the LQG and WDRC methods averaged over 1,000 simulation runs in the case of uniform disturbances. The shaded regions represent $25\%$ of the standard deviation.}
    \label{fig:cost_comp}
\end{figure}

Fig.~\ref{fig:cost_comp} showcases the effect of distributional uncertainties in  measurement noise. Specifically, it demonstrates the total costs incurred by the WDRC and LQG methods for measurement noise covariance matrix $M$ estimated using different samples. It is evident that even for only $10$ samples, the average performance of WDRC reaches that of LQG with fully known measurement noise distribution. These results illustrate the capabilities of our method to account for erroneous measurement noise information
although the proposed controller is designed to achieve distributional robustness in terms of disturbances. 
Using the worst-case distribution in the state estimator in our approach induces additional robustness to the Kalman filter, yielding better overall performance even for a small sample size compared to the standard LQG control method.

\section{Conclusions}
 In this work, we have presented a novel WDRC method for discrete-time partially observable linear systems. We have proposed an approximation scheme for reformulating the original WDRC problem into a tractable one. The approximate problem is first solved in finite-horizon settings, resulting in a closed-form expression of the optimal control policy with the corresponding Riccati equation. The mean vector of the worst-case distribution is also found in closed form, while the covariance matrix is found as the solution of a tractable SDP problem. The results for the finite-horizon problem were extended to the infinite-horizon setting by observing the asymptotic behaviors of the optimal policy pair and the cost. Consequently, we obtained a steady-state control policy by solving an ARE. The proposed method has several salient features, such as guaranteed cost property, probabilistic out-of-sample performance guarantee, and closed-loop stability. The experiment results demonstrate the capabilities of our method to immunize  partially observable linear systems against distributional ambiguity.
 
In the future, we plan to extend the obtained results to the case where the probability distribution of measurement noise  is also unknown. Therefore, the optimal controller must be robust against uncertainties in the distributions of both system disturbance and measurement noise. Moreover, to improve the tractability of the proposed method, it is worth examining conditions under which the SDP problem has a closed-form solution.

\appendix
\section{Intractability of Minimax LQ Control Problems with Wasserstein Penalty under Partial Observations}\label{app:po}
Consider the partially observable system~\eqref{sys} and the corresponding minimax control problem~\eqref{penalty0} in a finite horizon:
\[
\min_{\pi\in\Pi}\max_{\gamma\in\Gamma} \tilde{J}_T^\lambda (\pi, \gamma),
\]
where
\[
\tilde{J}_{T}^\lambda(\pi,\gamma) = \mathbb{E}_\mathbf{y}\bigg[\mathbb{E}_{x_{T}}[x_T^\top Q_f x_T\mid I_{T}] +\sum_{t=0}^{T-1} \mathbb{E}_{x_t}[x_t^\top Q x_t + u_t^\top R u_t \mid I_t, u_t] - \lambda W_2 (\mathbb{P}_t, \mathbb{Q}_t)^2\bigg].
\]
To solve the minimax control problem using DP, we define the value function recursively as follows:
\begin{equation}\label{vf_po}
\tilde{V}_t  (I_t) :=  \inf_{u_t\in\mathbb{R}^{n_u}} \sup_{\mathbb{P}_t\in\mathcal{P}(\mathbb{R}^{n_x})} \mathbb{E}_{x_t} [x_t^\top Q x_t  + u_t^\top R u_t -\lambda W_2(\mathbb{P}_t,\mathbb{Q}_t)^2 + \mathbb{E}_{y_{t+1}}[\tilde{V}_{t+1} (I_t, y_{t+1}, u_t) \mid I_t, u_t]
\end{equation}
with $\tilde{V}_T(I_T):=\mathbb{E}_{x_T}[x_T^\top Q_f x_T \mid I_T]$.

In fully observable settings, a common approach to solving the inner maximization problem in~\eqref{vf_po} is to use  Kantorovich duality~\cite{gao2016distributionally}.
The most tractable case is when 
the nominal distribution $\mathbb{Q}_t$ is chosen as the empirical distribution~\eqref{eq:emp}. 
In this case, Kantorovich duality can be expressed as
\begin{equation}\label{kant}
\sup_{\mathbb{P} \in \mathcal{P}(\mathbb{R}^{n_x})} \mathbb{E}_w [f(x, w)] - \lambda W_2(\mathbb{P},\mathbb{Q})^2  = \frac{1}{N} \sum_{i=1}^{N} \sup_{w \in \mathbb{R}^{n_x}} \big\{ f(x, w) - \lambda \|\hat{w}^{(i)} - w\|^2\big\}, 
\end{equation}
where $f:\mathbb{R}^{n_x}\times \mathbb{R}^{n_x} \to \mathbb{R}$ is some function depending on the disturbance $w$ and some fixed parameters $x$.

However, unlike the fully observable case, the uncertainty of the system is represented by the output $y_{t+1}$ and not $w_t$ directly. Therefore, if we can write the value function~\eqref{vf_po} in a way that has the form of the left-hand side in~\eqref{kant}, then  Kantorovich duality can be applied analogously to the fully observable settings. 
To this end, we recursively solve~\eqref{vf_po} to check whether a specific form of the value function is preserved. For time $t=T-1$, the value function is given by
\begin{equation*}
\begin{split}
\tilde{V}_{T-1} (I_{T-1}) =   \, & \inf_{u_{T-1}\in\mathbb{R}^{n_u}} \mathbb{E}_{x_{T-1}} [x_{T-1}^\top Q x_{T-1} \mid I_{T-1}]+ u_{T-1}^\top R u_{T-1} \\
& + \sup_{\mathbb{P}_{T-1}\in\mathcal{P}(\mathbb{R}^{n_x})}  \mathbb{E}_{x_{T-1}, w_{T-1}}[(A x_{T-1} + B u_{T-1} + w_{T-1})^\top \\
&\times Q_f (A x_{T-1} + B u_{T-1} + w_{T-1}) \mid I_{T-1}, u_{T-1}] -\lambda W_2(\mathbb{P}_{T-1},\mathbb{Q}_{T-1})^2.
\end{split}
\end{equation*}
It follows from Kantorovich duality that
\[
\begin{split}
\tilde{V}_{T-1}(I_{T-1}) = \, & \inf_{u_{T-1}\in\mathbb{R}^{n_u}} \mathbb{E}_{x_{T-1}} [x_{T-1}^\top Q x_{T-1} \mid I_{T-1}] + u_{T-1}^\top R u_{T-1} \\
& + \frac{1}{N} \sum_{i=1}^{N} \sup_{w_{T-1} \in \mathbb{R}^{n_x}}\big\{ \mathbb{E}_{x_{T-1}}(A x_{T-1}+ B u_{T-1}  + w_{T-1})^\top \\
& \times Q_f (A x_{T-1} + B u_{T-1} + w_{T-1}) \mid I_{T-1}, u_{T-1}] - \lambda \|\hat{w}_{T-1}^{(i)} - w_{T-1}\|^2 \big\}.
\end{split}
\]

If the penalty parameter satisfies the condition $\lambda I \succ Q_f$, then the inner maximization problem for each $i=1,\dots,N$ has a unique maximizer $w^{(i),*}$, given by
\[
w_{T-1}^{(i),*} := (\lambda I - Q_f)^{-1}\big[Q_f (A \mathbb{E}_{x_{T-1}}[x_{T-1}\mid I_{T-1}] + B u_{T-1}) + \lambda \hat{w}_{T-1}^{(i)}\big].
\]
Solving the outer minimization problem with respect to $u_{T-1}$ yields the following unique minimizer:
\[
u_{T-1}^* = -R^{-1} B^\top (I+Q_f B R^{-1} B^\top - \frac{1}{\lambda} Q_f)^{-1} \times \big(A \mathbb{E}_{x_{T-1}}[x_{T-1}\mid I_{T-1}] + \frac{1}{N}\sum_{i=1}^{N}\hat{w}_{T-1}^{(i)}\big).
\]
Then, the value function at time $t=T-1$ has the following quadratic form:
\[
\tilde{V}_{T-1} = \mathbb{E}_{x_{T-1}}[x_{T-1}^\top P_{T-1}  x_{T-1} + \xi_{T-1}^\top S_{T-1} \xi_{T-1} + 2 r_{T-1}^\top x_{T-1} \mid I_{T-1}] + q_{T-1},
\]
where $\xi_{T-1} = x_{T-1} - \mathbb{E}_{x_{T-1}}[x_{T-1}\mid I_{T-1}]$ is the difference between the state and its estimate, while $P_{T-1}, S_{T-1} \in \mathbb{S}_+^{n_x}, r_{T-1}\in\mathbb{R}^{n_x}$ and $q_{T-1}\in\mathbb{R}$ are coefficients.

Continuing the recursion for $t=T-2$, the  value function is written as
\begin{equation*}
\begin{split}
\tilde{V}_{T-2}(I_{T-2}) = \, & \inf_{u_{T-2}\in\mathbb{R}^{n_u}} \mathbb{E}_{x_{T-2}} [x_{T-2}^\top Q x_{T-2} \mid I_{T-2}] +  u_{T-2}^\top R u_{T-2}\\
& + \sup_{\mathbb{P}_{T-2}\in\mathcal{P}(\mathbb{R}^{n_x})} \mathbb{E}_{x_{T-2}, w_{T-2}}[(A x_{T-2} + B u_{T-2} + w_{T-2})^\top P_{T-1} (A x_{T-2} + B u_{T-2} + w_{T-2})\\
& + 2r_{T-1}^\top (A x_{T-2} + B u_{T-2} + w_{T-2}) \mid I_{T-2}, u_{T-2}] + \mathbb{E}_{y_{T-1}, x_{T-1}}[\xi_{T-1}^\top S_{T-1} \xi_{T-1}\mid I_{T-1}]\\
& + q_{T-1} -\lambda W_2(\mathbb{P}_{T-2},\mathbb{Q}_{T-2})^2 .
\end{split}
\end{equation*}
Due to the structure of the expression inside the maximization, it is straightforward that the value function does not have the form in~\eqref{kant}. This is because $\mathbb{E}_{y_{T-1},x_{T-1}}[\xi_{T-1}^\top S_{T-1} \xi_{T-1}\mid I_{T-1}]$ cannot be represented by an expectation with respect to $w_{T-2}$, though it implicitly depends on the disturbances via $x_{T-1}$ and $y_{T-1}$. Consequently,
the standard LQR argument is not applicable to
 the minimax problem with the Wasserstein penalty under partial observations.

\section{Proofs}

\subsection{Proof of Lemma~\ref{lem:opt_cont}}\label{app:opt_cont}
\begin{proof}
Having the quadratic value function for time $t+1$ and plugging it  into~\eqref{bellman}, the value function for time $t$ is given by
\[
\begin{split}
V_t( I_t)  = \,&  \inf_{u_t\in\mathbb{R}^{n_u}} \sup_{\substack{\bar{w}_t \in\mathbb{R}^{n_x},\\ \Sigma_t \in\mathbb{S}_+^{n_x}}}\mathbb{E}_{x_t}[x_t^\top Q x_t\mid I_t] + u_t^\top R u_t   + \mathbb{E}_{x_t,w_t}\big[(A x_t + Bu_t + w_t)^\top P_{t+1}(A x_t + Bu_t + w_t)\\
&+ 2r_{t+1}^\top (A x_t + Bu_t + w_t)  \mid I_t, u_t\big] + \mathbb{E}_{x_{t+1},y_{t+1}}[\xi_{t+1}^\top S_{t+1}\xi_{t+1} \mid I_{t}]\\
&+ q_{t+1} -\lambda [ \| \bar{w}_t - \hat{w}_t \|^2 + \mathrm{B}^2(\Sigma_t, \hat{\Sigma}_t ) ].
\end{split}
\]
Using the property that 
\[
\mathbb{E}[w_t^\top P_{t+1} w_t] = \bar{w}_t^\top P_{t+1} \bar{w}_t + \mathrm{Tr}[P_{t+1}\Sigma_t],
\]
we further simplify the value function as
\[
\begin{split}
V_t( I_t)  = \, & \inf_{u_t\in\mathbb{R}^{n_u}}\sup_{\substack{\bar{w}_t \in\mathbb{R}^{n_x},\\ \Sigma_t \in\mathbb{S}_+^{n_x}}}\mathbb{E}_{x_t}[x_t^\top Q x_t\mid I_t] + u_t^\top R u_t   + \mathbb{E}_{x_t}\big[(A x_t + Bu_t +\bar{w}_t)^\top P_{t+1}(A x_t + Bu_t + \bar{w}_t)\\
&+ 2r_{t+1}^\top (A x_t + Bu_t + \bar{w}_t)  \mid I_t, u_t\big]- \lambda \|\bar{w}_t - \hat{w}_t\|_2^2  + \mathbb{E}_{x_{t+1},y_{t+1}}[\xi_{t+1}^\top S_{t+1}\xi_{t+1} \mid I_{t}]\\
&+ \mathrm{Tr}[(P_{t+1}-\lambda I)\Sigma_{t}]+ 2\lambda\mathrm{Tr}[(\hat{\Sigma}_t^{1/2}\Sigma_t \hat{\Sigma}_t^{1/2})^{1/2}]- \lambda\mathrm{Tr}[\hat{\Sigma}_{t}] + q_{t+1}.
\end{split}
\]
Note that
\[
\mathbb{E}_{x_{t+1},y_{t+1}}[\xi_{t+1}\mid I_t] = 0,
\]
and $\mathbb{E}_{x_{t+1}, y_{t+1}}[\xi_{t+1}\xi_{t+1}^\top \mid I_t]$ is independent of $u_t$ and $\bar{w}_t$.
Thus,  the objective function for the inner maximization problem
\[
\begin{split}
\mathbb{E}_{x_t}\big[(A x_t + Bu_t +\bar{w}_t)^\top & P_{t+1}(A x_t + Bu_t + \bar{w}_t)+ 2r_{t+1}^\top (A x_t + Bu_t + \bar{w}_t)  \mid I_t, u_t\big]- \lambda \|\bar{w}_t - \hat{w}_t\|_2^2 \\
&+ \mathbb{E}_{x_{t+1},y_{t+1}}[\xi_{t+1}^\top S_{t+1}\xi_{t+1} \mid I_{t}] + \mathrm{Tr}[(P_{t+1} - \lambda I)\Sigma_t]+ 2\lambda\mathrm{Tr}[(\hat{\Sigma}_t^{1/2}\Sigma_t \hat{\Sigma}_t^{1/2})^{1/2}]
\end{split}
\]
can be written separately in terms of $\bar{w}_t$ 
and $\Sigma_t$, enabling to solve two independent maximization problems. 
Specifically, the two problems are as follows:
\[
\max_{\bar{w}_t \in \mathbb{R}^{n_x}}  \mathbb{E}_{x_t}\big[(A x_t + Bu_t +\bar{w}_t)^\top P_{t+1}(A x_t + Bu_t + \bar{w}_t)+ 2r_{t+1}^\top (A x_t + Bu_t + \bar{w}_t)  \mid I_t, u_t\big]- \lambda \|\bar{w}_t - \hat{w}_t\|_2^2 
\]
and
\begin{equation}\label{cov_obj}
  \max_{\Sigma_t \in\mathbb{S}_+^{n_x}} \mathbb{E}_{x_{t+1},y_{t+1}}[\xi_{t+1}^\top S_{t+1} \xi_{t+1}\mid I_t] + \mathrm{Tr}[(P_{t+1} - \lambda I)\Sigma_t + 2\lambda (\hat{\Sigma}_t^{1/2}\Sigma_t \hat{\Sigma}_t^{1/2})^{1/2}].
\end{equation}

Regarding the first problem for $\bar{w}_t$, 
the Hessian of value function with respect to $\bar{w}_t$ is negative definite under the assumption on the penalty parameter $\lambda$.
Thus, 
the objective is strictly concave, and its  unique maximizer given control input $u_t$ is obtained
from the first-order optimality condition
 as
\begin{equation}\label{wbar_u}
\bar{w}_t^*(u_t) = (\lambda I - P_{t+1})^{-1} \big(P_{t+1}[A_t \bar{x}_t + B_t u_t] + r_{t+1}\big).
\end{equation}

Note that the maximizer of the second problem is independent of the control input $u_t$. 
For the outer minimization problem with respect to $u_t$,
we first differentiate the objective function with respect to $u_t \in \mathbb{R}^{n_u}$ to obtain the following derivative:
\begin{equation}\label{opt_cond}
\begin{split}
& 2\left[B + \frac{\partial \bar{w}^*_t(u_t)}{\partial u_t}\right]^\top [P_{t+1}(A\bar{x}_t + B u_t + \bar{w}_t^*(u_t)) + r_{t+1}] - 2\lambda \frac{\partial\bar{w}_t^*(u_t)}{\partial u_t}(\bar{w}_t^*(u_t) - \hat{w}_t) + 2R u_t\\
&=   2 B^\top g_t(u_t) + 2 R u_t,
\end{split}
\end{equation}
where
\begin{equation}\label{g_u}
g_t(u_t) := P_{t+1}(A\bar{x}_t + B u_t + \bar{w}_t^* (u_t)) + r_{t+1}.
\end{equation}
Differentiating the derivative with respect to $u_t$ again, we can check that the Hessian of the objective function is positive definite under the assumption on the penalty parameter $\lambda$. 
Thus, the unique minimizer $u_t^*$ can be obtained by using the first-order optimality condition:
\begin{equation}\label{cont}
    u_t^* = - R^{-1} B^\top g_t^*.
\end{equation}

For further simplifications, we let $\bar{w}_t^*=\bar{w}_t^*(u_t^*)$ and rewrite it as
\[
\bar{w}_t^* = \frac{1}{\lambda}\big(P_{t+1}(A\bar{x}_t + Bu_t^* +\bar{w}_t^*) + r_{t+1} + \lambda\hat{w}_t\big),
\]
which yields the following expression for $g_t^*$:
\[
g_t^* = P_{t+1}\left(A\bar{x}_t - B R^{-1}B^\top g_t^* + \frac{1}{\lambda}g_t^* + \hat{w}_t\right) + r_{t+1}.
\]
Finally, we have
\begin{equation}\label{g_star}
g_t^* = (I + P_{t+1}\Phi)^{-1} (P_{t+1} A\bar{x}_t + P_{t+1} \hat{w}_t + r_{t+1})
\end{equation}
and
\begin{equation}\label{w_star}
\bar{w}_t^* = \frac{1}{\lambda} g_t^* + \hat{w}_t.
\end{equation}
We conclude the proof by replacing~\eqref{g_star} into~\eqref{cont}.
\end{proof}

\subsection{Proof of Theorem~\ref{thm:sol}}\label{app:sol}
\begin{proof}
We use mathematical induction backward in time to prove the theorem. For $t = T$, by definition,
the value function is in the desired form 
\[
V_T(I_T) = \mathbb{E}_{x_T}[x_T^\top P_T x_T | I_{T}].
\]

Now, it suffices to show that $V_{t}$ is in the required form, given that $V_{t+1}$ is in that form. Specifically, the value function at time $t$ can be written as
\[
\begin{split}
V_t( I_t)  = \, & \inf_{u_t\in\mathbb{R}^{n_u}} \sup_{\substack{\bar{w}_t \in\mathbb{R}^{n_x},\\ \Sigma_t \in\mathbb{S}_+^{n_x}}}\mathbb{E}_{x_t}[x_t^\top Q x_t\mid I_t] + u_t^\top R u_t   + \mathbb{E}_{x_t,w_t}\big[(A x_t + Bu_t + w_t)^\top P_{t+1}(A x_t + Bu_t + w_t)\\
&+ 2r_{t+1}^\top (A x_t + Bu_t + w_t)  \mid I_t, u_t\big] + \mathbb{E}_{x_{t+1},y_{t+1}}[\xi_{t+1}^\top S_{t+1}\xi_{t+1} \mid I_{t}] -\lambda [ \| \bar{w}_t - \hat{w}_t \|^2 + \mathrm{B}^2(\Sigma_t, \hat{\Sigma}_t ) ] \\
&+ q_{t+1} + \sum_{s=t+1}^{T-1}\mathbb{E}_{y_{t+1}}[z_t(I_{t}, u_t, y_{t+1}, s) \mid I_t ,u_t].
\end{split}
\]

It follows from the law of total expectation that
\[
\mathbb{E}_{y_{t+1}}[z_{t+1}(I_{t}, u_t, y_{t+1},s)\mid I_t,u_t] =  z_t(I_t, s),
\]
which is independent of $\bar{w}_t$, $\Sigma_t$, and $u_t$. Therefore, by Lemma~\ref{lem:opt_cont}, the mean vector~\eqref{mu} and the covariance matrix solving~\eqref{max_sigma} are an optimal solution pair of the inner maximization problem. Moreover, the optimal value of~\eqref{max_sigma} corresponds to $z_t(I_t,t)$. Meanwhile, the outer minimization problem has a unique optimal solution given as~\eqref{u_opt}. By plugging these values into the Bellman equation, we have
\[
\begin{split}
V_t( I_t)  = \, & \mathbb{E}_{x_t}[x_t^\top Q x_t\mid I_t] + (g_t^*)^\top B R^{-1} B^\top g_t^* + \mathbb{E}_{x_t}\big[(A x_t - \Phi g_t^* + \hat{w}_t)^\top P_{t+1}(A x_t - \Phi g_t^* + \hat{w}_t)\\
&+ 2r_{t+1}^\top (A x_t - \Phi g_t^* + \hat{w}_t)  \mid I_t, u_t\big]- \frac{1}{\lambda} (g_t^*)^\top g_t^*  - \lambda\mathrm{Tr}[\hat{\Sigma}_{t}] + q_{t+1} + z_t(I_t,t) + \sum_{s=t+1}^{T-1}z_t(I_t,s).
\end{split}
\]

It remains to simplify the expression by substituting the values for $r_t$ and $q_t$ as in~\eqref{r} and~\eqref{q}. Then, the value function for time $t$ can be written as
\[
V_t(I_t)  = \mathbb{E}_{x_t}[x_t^\top (Q + A^\top P_{t+1} A) x_t \mid I_t]- \bar{x}_{t}^\top S_t \bar{x}_t + 2r_t^\top \bar{x}_t + q_{t}  + \sum_{s=t}^{T-1}z_t(I_{t},s),
\]
where $S_t = A^\top P_{t+1} \Phi(I+P_{t+1}\Phi)^{-1} P_{t+1} A$. This can be expressed as
\[
\begin{split}
V_t(I_t)  & = \mathbb{E}_{x_t}[x_t^\top (Q + A^\top P_{t+1} A - S_{t}) x_t\mid I_t] + \mathbb{E}_{x_t}  [\xi_t^\top S_t\xi_t + 2r_t^\top x_t\mid I_t]  + q_{t} + \sum_{s=t}^{T-1}z_{t}(I_t, s)\\
&= \mathbb{E}_{x_t}[x_t^\top P_t x_t + \xi_t^\top S_{t}\xi_t + 2 r_t^\top x_t \mid I_t] + q_{t} + \sum_{s=t}^{T-1}z_{t}(I_t, s),
\end{split}
\]
which is in the desired form with parameters~\eqref{P}--\eqref{z}. This completes our inductive argument.

So far, we have shown that the value function is measurable, and the outer minimization
 problem in the Bellman equation~\eqref{bellman} admits an optimal solution.
 Thus, 
  it follows from the DP principle that the control policy $\pi^*$ constructed as that in the theorem statement is optimal.
  Moreover, if \eqref{z} admits an optimal solution $\Sigma_t^*$ for all $t$, the policy pair $(\pi_t^*, \gamma_t^*)$ is minimax optimal. 
\end{proof}

\subsection{Proof of Proposition~\ref{prop:sdp}}\label{app:sdp}

First, we notice that
$\bar{X}_{t+1} = \mathbb{E}_{x_{t+1},y_{t+1}}[\xi_{t+1} \xi_{t+1}^\top \mid I_t]$.
It follows from the Kalman filter recursion~\eqref{kalman_cov} and~\eqref{cov_update} that $z_t(I_t, t)$ is equal to the optimal value of~\eqref{max_sigma}, which in its turn is equivalent to the following optimization problem:
\[
\begin{split}
\max_{\substack{X, X^{-},\\ \Sigma \in\mathbb{S}_+^{n_x} }} \; & \mathrm{Tr}[S_{t+1} X + (P_{t+1} - \lambda I)\Sigma+ 2\lambda (\hat{\Sigma}_t^{1/2} \Sigma \hat{\Sigma}_t^{1/2})^{1/2}]\\
\mbox{s.t.} \; & X = X^{-} - X^{-} C^\top (C X^{-} C^\top + M)^{-1} C X^{-}\\
& X^{-} = A \bar{X}_t A^\top + \Sigma,
\end{split}
\] 
where $\bar{X}_t$ is the state covariance matrix conditioned on the information vector $I_t$. The objective function here is continuous and jointly concave in $\Sigma, X^{-}$ and $X$ due to the positive semidefiniteness of $S_{t+1}$ and Assumption~\ref{ass:lambda_ass}. Therefore, the problem has an optimal solution and we can obtain
 optimal  $(\Sigma^*, X^*)$, corresponding to $\Sigma_t^*$ and $\bar{X}_{t+1}$.
The reformulation into the SDP form~\eqref{sdp} is performed by using the property that $\mathrm{Tr}[S_{t+1}X] \leq \mathrm{Tr}[S_{t+1}X']$ for any $X\preceq X'$ and then applying the Schur complement lemma to replace the inequality constraints with the corresponding linear matrix inequalities. 
 
\subsection{Proof of Proposition~\ref{prop:P_ss}}\label{app:P_ss}
\begin{proof}
The proof follows from the asymptotic property of the Riccati equation for the standard LQ control. Specifically, we rewrite the Riccati equation~\eqref{P} as follows:
\begin{equation}\label{P_perf}
\begin{split}
    P_t &= Q + A^\top (I + P_{t+1} \Phi^{1/2} I^{-1} (\Phi^{1/2})^\top )^{-1} P_{t+1} Q\\
    & = Q + A^\top (P_{t+1}  - P_{t+1} \tilde{B} (\tilde{R} + \tilde{B}^\top P_{t+1} \tilde{B})^{-1} \tilde{B}^\top P_{t+1}) A,
\end{split}
\end{equation}
where $\tilde{R} = I$, $\tilde{B} = \Phi^{1/2}$. Consider a hypothetical linear system $(A,\tilde{B})$ with a quadratic cost function replacing $R$ with $\tilde{R}$. It is evident that~\eqref{P_perf} has the form of the standard Riccati equation for this hypothetical LQ control problem. 
It follows from the standard LQ control theory that if the pair $(A,\tilde{B})$ is stabilizable and $(A,Q^{1/2})$ is detectable, then there exists a  $P_{ss}\succeq 0$ such that~\eqref{conv} holds for any $P_T \succeq 0$. Furthermore, it is the unique solution of the ARE~\eqref{are}~\cite[Proposition~3.1.1]{bertsekas2012dynamic}.
\end{proof}

\subsection{Proof of Lemma~\ref{lem:rs_ss}}\label{app:rs_ss}
\begin{proof}
It follows from Proposition~\ref{prop:P_ss} that $P_t \to P_{ss}$ as $T \to \infty$, and thus the convergence of $\{S_t\}$ to $S_{ss}$ is straightforward. Moreover, $r_t$ is updated according to
\[
 r_{t} =  A^\top (I+P_{ss} \Phi)^{-1}(r_{t+1} + P_{ss}\hat{w})
\]
as $T \to \infty$. Thus, to ensure the convergence of  $\{r_t\}$,
it suffices to show that $A^\top (I+P_{ss} \Phi)^{-1}$.
For this, we revisit the proof of Proposition~\ref{prop:P_ss}
and notice that the ARE can be expressed as
\[
P_{ss} = Q + A^\top (P_{ss} - P_{ss} \tilde{B} (\tilde{R} + \tilde{B}^\top P_{ss} \tilde{B})^{-1} \tilde{B}^\top P_{ss}) A,
\]
where $\tilde{R} = I$ and $\tilde{B} = \Phi^{1/2}$.  
Then, the optimal control gain matrix for the hypothetical LQ control problem for the linear system $(A,\tilde{B})$ with a quadratic cost function replacing $R$ with $\tilde{R}$ is given by
\[
\tilde{K} = (\tilde{R} + \tilde{B}^\top P_{ss} \tilde{B})^{-1} \tilde{B}^\top P_{ss} A,
\]
and the closed-loop ``A" matrix is
\[
A + \tilde{B} \tilde{K} = A - \tilde{B} (\tilde{R} + \tilde{B}^\top P_{ss} \tilde{B})^{-1} \tilde{B}^\top P_{ss} A,
\]
which is stable because $(A, \tilde{B})$ is stabilizable. 
Since $A^\top (I+P_{ss} \Phi)^{-1} = (A + \tilde{B} \tilde{K} )^\top$, it is also a stable matrix. 
Therefore, $\{r_t\}$ converges to its limit, which is obtained as~\eqref{r_ss}. 
\end{proof}

\subsection{Proof of Proposition~\ref{prop:avg_cost}}\label{app:avg_cost}
\begin{proof}
It follows from Theorem~\ref{thm:sol} that the finite-horizon cost incurred by the policy pair $(\pi_{ss}^*,\gamma_{ss}^*)$ is given by
\[
\begin{split}
J_T^\lambda (\pi_{ss}^*,\gamma_{ss}^*)= \, & \mathbb{E}_{y_0}\big[\mathbb{E}_{x_0} [ x_0^\top P_{0} x_0 + \xi_0^\top S_{0} \xi_0 + 2 r_{0}^\top x_0 \mid I_0]\big] + q_{0} \\
&+ \sum_{t=0}^{T-1}\Big(\mathrm{Tr}[S_{t+1} \bar{X}_{t+1}+(P_{t+1} - \lambda I) \Sigma_{ss}^*] +2\lambda \mathrm{Tr}[ (\hat{\Sigma}^{1/2} \Sigma_{ss}^*\hat{\Sigma}^{1/2})^{1/2}]\Big),
\end{split}
\]
where $\bar{X}_{t+1}$ is the state covariance matrix computed using $\Sigma_{ss}^*$. It follows from~\eqref{ss_cov} that $\{\bar{X}_{t+1}\}$ converges to $\bar{X}_{ss}$ as $T\to\infty$. By the convergence of $P_t, S_t$, and $r_t$, as well as the recursion for $q_t$, the steady-state average cost is given by
\[
\begin{split}
\rho  = \,& \limsup_{T\to\infty} \frac{1}{T}J_T^\lambda (\pi_{ss}^*,\gamma_{ss}^*)
\\
 =\, &\mathrm{Tr}[S_{ss} \bar{X}_{ss} +(P_{ss} - \lambda I) \Sigma_{ss}^*+2\lambda (\hat{\Sigma}^{1/2} \Sigma_{ss}^*\hat{\Sigma}^{1/2})^{1/2}]\\
& +(2\hat{w}  -  \Phi r_{ss})^\top  ( I+P_{ss}\Phi)^{-1}r_{ss} - \lambda\mathrm{Tr}[\hat{\Sigma}] +\hat{w}^\top (I+P_{ss}\Phi)^{-1}P_{ss}\hat{w}.
\end{split}
\]
The first term in the last equation corresponds to the optimal value $z_{ss}$ of the maximization problem~\eqref{z_ss_opt}. Therefore, the result follows.
\end{proof}

\subsection{Proof of Proposition~\ref{prop:bellman}}\label{app:bellman}
 \begin{proof}
We first rewrite $h$ as
\[
h(I_t) = \mathbb{E}_{x_t}[x_t^\top P_{ss} x_t + \xi_t^\top S_{ss} \xi_t + 2 r_{ss}^\top x_t \mid I_t]
\]
with $\mathbb{E}_{x_{t}}[\xi_{t} \xi_{t}^\top \mid I_{t}] = X_{t} \equiv \bar{X}_{ss}$.
Next, we apply Lemma~\ref{lem:opt_cont} by letting $V_{t+1} \equiv h$, or, by setting $P_{t+1} = P_{ss}, S_{t+1} = S_{ss}, r_{t+1} = r_{ss}$, and $q_{t+1} = 0$.
Then, the minimax problem on the right-hand side of~\eqref{bellman_1} has the optimal value of
\[
\mathbb{E}_{x_t}[x_t^\top P_t x_t + \xi_t^\top S_t \xi_t + 2 r_t^\top x_t\mid I_t] + q_t + z_t(I_t, t),
 \]
 where
 \begin{align*}
     P_t & = Q + A^\top (I + P_{ss}\Phi)^{-1} P_{ss} A \\
     S_t & =  Q + A^\top  P_{ss} A - P_{ss}\\
     r_t & =  A^\top (I + P_{ss}\Phi)^{-1} ( r_{ss} + P_{ss}\hat{w})\\
     q_t & = (2\hat{w} - \Phi r_{ss})^\top(I + P_{ss}\Phi)^{-1} r_{ss}+\hat{w}^\top (I + P_{ss}\Phi)^{-1}  P_{ss} \hat{w} - \lambda\mathrm{Tr}[\hat{\Sigma}],
 \end{align*}
 and
 \begin{equation}\label{z_aux}
 z_t(I_t,t) =   \sup_{\Sigma_t\in\mathbb{S}^{n_x}_+} \mathrm{Tr}[S_{ss} \bar{X}_{t+1}] + \mathrm{Tr}[(P_{ss}-\lambda I)\Sigma_t + 2\lambda (\hat{\Sigma}^{1/2}\Sigma_t\hat{\Sigma}^{1/2})].
 \end{equation}
 
It follows from the ARE~\eqref{are} that $P_t = P_{ss}$, while from~\eqref{S_ss} and~\eqref{r_ss} we have  $S_t = S_{ss}$ and $r_t = r_{ss}$, respectively. Since $\bar{X}_{t+1} = \bar{X}_{ss}$ is stationary, the maximization problem~\eqref{z_ss_opt} yields $z_t(I_t,t) = z_{ss}$ with its maximizer corresponding to the stationary covariance matrix $\Sigma_{ss}^*$.
Moreover, we have
\begin{align*}
    \bar{X}_{ss} &= \bar{X}_{t+1}^{-} - \bar{X}_{t+1}^{-} C^\top(C \bar{X}_{t+1}^{-} C^\top + M)^{-1}C \bar{X}_{t+1}^{-}\\
    \bar{X}_{t+1}^{-} &= A \bar{X}_{t} A^\top + \Sigma_{ss}^*,
\end{align*}
which is valid only if $\bar{X}_{t}=\bar{X}_{ss}$. As a result, the optimal value of the minimax problem is equal to
 \[
\bar{x}_t^\top P_{ss} \bar{x}_t + 2 r_{ss}^\top \bar{x}_t + \mathrm{Tr}[(S_{ss} + P_{ss}) \bar{X}_{ss}] + q_{t} + z_{ss}.
 \]
Thus,  the equality in~\eqref{bellman_1} holds. The optimality of the solution pair $(\pi_{ss}^*(I_t), \gamma_{ss}^*(I_t))$ follows directly from Lemma~\ref{lem:opt_cont}.
 \end{proof}

\subsection{Proof of Proposition~\ref{prop:minmax_opt}}\label{app:minmax_opt}
\begin{proof}
Fix an arbitrary control policy $\pi:=(\pi_0,\pi_1,\dots) \in \Pi$. 
We first show that
\begin{equation}\label{ind_hyp}
\bar{J}_T^\lambda(\pi,\gamma_{ss}^*) \geq T\rho + \mathbb{E}_{y_0}[h(I_0)]
\end{equation}
using mathematical induction. 
For $T = 0$, $\bar{J}_0^\lambda (\pi, \gamma_{ss}^*) = \mathbb{E}_{y_0} [ h (I_0)]$.
Suppose that the induction hypothesis is true for $T = k$. 
When $T = k+1$, it follows from Proposition~\ref{prop:bellman} that
\begin{equation}\nonumber
\begin{split}
\bar{J}_{k+1}^\lambda (\pi, \gamma_{ss}^*)
&\geq \bar{J}_{k}^\lambda (\pi, \gamma_{ss}^*)
-\mathbb{E}_{y_{0:k}}[h(I_k)] + \rho + \mathbb{E}_{y_{0:k}}[ h(I_k)]\\
&\geq (k+1) \rho + \mathbb{E}_{y_0}[h(I_0)]. 
\end{split}
\end{equation}
This completes our inductive argument. 

Dividing both sides of \eqref{ind_hyp} by $T$ and taking $\limsup$, we obtain that
\begin{equation}\label{one_sided_1}
\bar{J}_{\infty}^\lambda (\pi,\gamma_{ss}^*) \geq \rho,
\end{equation}
which holds for any control policy $\pi\in\Pi$. 

Now, for any $\pi \in \bar{\Pi}$, 
the left-hand side of \eqref{one_sided_1} is equivalent to
\begin{equation}\label{equal}
\begin{split}
\bar{J}_\infty^\lambda(\pi, \gamma_{ss}^*) =  \, &\limsup\limits_{T\to\infty}\frac{1}{T}\mathbb{E}_\mathbf{y}[h(I_T)\mid \pi, \gamma_{ss}^* ]  \\
&+ \limsup\limits_{T\to\infty}\frac{1}{T}\mathbb{E}_\mathbf{y}\bigg[\sum_{t=0}^{T-1} \mathbb{E}_{x_t}[x_t^\top Q x_t\mid I_t] + u_t^\top R u_t - \lambda \mathrm{G} (\mathbb{P}_t, \mathbb{Q}_t)^2 \mid \pi, \gamma_{ss}^*\bigg] \\ 
= \, &J_\infty^\lambda(\pi,\gamma_{ss}^*),
\end{split}
\end{equation}
with the last equality following from the condition~\eqref{cond_1}.
Combining~\eqref{one_sided_1} and~\eqref{equal} yields
\[
J_\infty^\lambda(\pi,\gamma_{ss}^* )\geq \rho \quad \forall \pi \in \bar{\Pi}.
\]
Using a similar argument, we can show that
\[
J_\infty^\lambda(\pi_{ss}^*,\gamma )\leq \rho \quad \forall \gamma \in \bar{\Gamma}.
\]
Therefore, $(\pi_{ss}^*, \gamma_{ss}^*)$ is minimax optimal, and the optimal value corresponds to $\rho$.
%
\end{proof}

\subsection{Proof of Proposition~\ref{prop:ellipt}}\label{app:ellipt}
\begin{proof}
Since $\gamma^* \in \Gamma$, it is admissible to the original minimax control problem~\eqref{penalty0}. Also, by Lemma~\ref{lem:gelbrich}, if the nominal distribution $\mathbb{Q}_t$ is elliptical, then~\eqref{bound} holds with equality, yielding
\begin{equation}\nonumber
J^\lambda_\infty (\pi, \gamma^*) = \tilde{J}^\lambda_\infty (\pi, \gamma^*) \quad \forall \pi \in \Pi. 
\end{equation}
Therefore, 
\[
J^\lambda_\infty (\pi^*, \gamma^*)
= \inf_{\pi \in \Pi}J ^\lambda_\infty (\pi, \gamma^*) \leq \tilde{J}^\lambda_\infty (\pi, \gamma^*) \quad \forall \pi \in \Pi.
\]

On the other hand, Lemma~\ref{lem:gelbrich} implies that 
\begin{equation}\nonumber
\begin{split}
J^\lambda_\infty (\pi^*, \gamma^*)
&=
\sup_{\gamma\in\Gamma} J_\infty^\lambda (\pi^*, \gamma)\\
& \geq \sup_{\gamma\in\Gamma}\tilde{J}_\infty^\lambda (\pi^*, \gamma) \geq \tilde{J}_\infty^\lambda (\pi^*, \gamma) \quad  \forall \gamma \in \Gamma.
\end{split}
\end{equation}
Finally, we obtain that
\[
\tilde{J}_\infty^\lambda (\pi^*, \gamma) \leq
J^\lambda_\infty (\pi^*, \gamma^*) \leq \tilde{J}^\lambda_\infty (\pi, \gamma^*) \quad \forall (\pi, \gamma) \in \Pi \times \Gamma.
\]
This implies that $(\pi^*, \gamma^*)$ is minimax optimal to the original problem~\eqref{penalty0}.
\end{proof}

\subsection{Proof of Theorem~\ref{thm:guarantee}}\label{app:guarantee}

\begin{proof}
Fix $\lambda >0$. 
Let $\mathrm{LHS} := \sup_{\gamma \in \bar{\Gamma}_{\mathcal{D}}} J_\infty (\pi_{ss}^{\lambda, \star}, \gamma)$
and $\mathrm{RHS} := \theta^2 \lambda + \rho (\lambda)$. 
For any $\varepsilon > 0$, there exists $\gamma^\varepsilon \in \bar{\Gamma}_{\mathcal{D}}$ such that 
\[
\mathrm{LHS} - \epsilon < J_\infty (\pi_{ss}^{\lambda, \star}, \gamma^\varepsilon).
\]
By Lemma~\ref{lem:gelbrich} and the definition of the Wasserstein ambiguity set $\mathcal{D}_t$, we have 
\[
\mathrm{G}(\mathbb{P}_t,\mathbb{Q}_t)^2 \leq W_2(\mathbb{P}_t, \mathbb{Q}_t)^2 \leq \theta^2 \quad \forall \mathbb{P}_t\in \mathcal{D}_t.
\]
Thus, it follows from  $\gamma^\epsilon \in \bar{\Gamma}_{\mathcal{D}}$ and the definitions of $J_\infty$ and $J_\infty^\lambda$
 that
\begin{equation} \nonumber
\begin{split}
J_\infty (\pi_{ss}^{\lambda, \star}, \gamma^\varepsilon) 
&\leq \theta^2 \lambda   + J_\infty^\lambda (\pi_{ss}^{\lambda, \star}, \gamma^\varepsilon) \\
&\leq \theta^2 \lambda   + \sup_{\gamma \in \bar{\Gamma}} J_\infty^\lambda (\pi_{ss}^{\lambda, \star}, \gamma) =  \theta^2\lambda + \rho (\lambda).
\end{split}
\end{equation}
Since $\epsilon$ was arbitrarily chosen, 
$\mathrm{LHS} \leq \mathrm{RHS}$ as desired. 
\end{proof}

\subsection{Proof of Theorem~\ref{thm:perf_guar}}\label{app:perf_guar}
\begin{proof}
It follows from the measure concentration inequality~\eqref{measure} that for a Wasserstein ambiguity set with radius $\theta$ chosen according to~\eqref{radius}, the following probabilistic bound holds:
\begin{equation}\label{prob_1}
\mathbb{P}^{N}\{\bold{w} \mid W_2(\mathbb{P},\mathbb{Q}) \leq \theta\} \geq 1-\beta,
\end{equation}
meaning that the true distribution $\mathbb{P}$ lies in the ambiguity set with a probability no less than $(1-\beta)$. 

Moreover, Theorem~\ref{thm:guarantee} suggests
\[
J_\infty (\pi_{ss,\bold{w}}^{\lambda(\theta), *},\gamma) \leq  \theta^2  \lambda(\theta) + \rho(\lambda(\theta)) \quad \forall \gamma\in\bar{\Gamma}_\mathcal{D}.
\]
Finally, the true distribution $\mathbb{P}$ belongs to the ambiguity set $\mathcal{D}$ with a probability no less than $(1-\beta)$, the inequality holds with the same probability, thereby concluding the proof.
\end{proof}

\subsection{Proof of Proposition~\ref{prop:stability}}\label{app:stability}
\begin{proof}
The mean-state system under the optimal policy $(\pi_{ss}^*, \gamma_{ss}^*)$ can be written as
\begin{equation}\label{closed_loop_state}
\begin{split}
\tilde{x}_{t+1} = \, & A \tilde{x}_t + (B K_{ss} + H_{ss}) \bbar{x}_t + B L_{ss} + G_{ss}\\
\bbar{x}_{t+1} =\, & (A + B K_{ss} + H_{ss} - \bar{X}_{ss} C^\top M^{-1} C A) \bbar{x}_t  \\
&+ B L_{ss} + G_{ss} +\bar{X}_{ss} C^\top M^{-1} C A \tilde{x}_t.
\end{split}
\end{equation}

Let $e_t:=\tilde{x}_t - \bbar{x}_t$ be the \emph{error state}, representing the difference between the expected values of the true state and its estimate. Then, the error state evolves according to
\begin{equation*}
\begin{split}
    e_{t+1} & =  (A -  \bar{X}_{ss} C^\top M^{-1} C A) (\tilde{x}_t - \bbar{x}_t)\\
    & = (A -  \bar{X}_{ss}^{-} C^\top (C \bar{X}_{ss}^{-} C^\top + M)^{-1} C A) e_t,
    \end{split}
\end{equation*}
where the last equation follows from the identity
\[
\bar{X}_{ss}C^\top M^{-1} = \bar{X}_{ss}^{-} C^\top (C \bar{X}_{ss}^{-} C^\top + M)^{-1}.
\]
For the steady-state Kalman filter, it is known that under Assumption~\ref{ass:cont_obs} the PSD matrix $\bar{X}_{ss}^-$ solves the filter ARE~\eqref{cov_update_ss}. Therefore, the corresponding closed-loop gain matrix $A -  \bar{X}_{ss}^{-} C^\top (C \bar{X}_{ss}^{-} C^\top + M)^{-1} C A$ has eigenvalues strictly within the unit circle, yielding
\begin{equation}\label{e_conv}
\lim_{t\to\infty} e_t = 0.
\end{equation}

On the other hand, it follows from~\eqref{closed_loop_state} that
\begin{equation}\label{state_new}
\tilde{x}_{t+1} = (A  + B K_{ss} + H_{ss}) \tilde{x}_t - (B K_{ss} + H_{ss}) e_t + BL_{ss} + G_{ss}.
\end{equation}
To show the convergence of $\{\tilde{x}_{t+1}\}$, we rewrite $H_{ss}$ and $G_{ss}$ as
\[
\begin{split}
  H_{ss} &= \frac{1}{\lambda} (I+P_{ss} \Phi)^{-1} P_{ss} A\\
  G_{ss} &= \frac{1}{\lambda} (I+P_{ss} \Phi)^{-1}(P_{ss} \hat{w} + r_{ss}) + \hat{w}.
\end{split}
\]
Substituting the above expressions and those for $K_{ss}$ and $L_{ss}$ into~\eqref{state_new}, we obtain
\[
\tilde{x}_{t+1} = (I+\Phi P_{ss})^{-1} A \tilde{x}_t + \Phi(I+P_{ss}\Phi)^{-1}P_{ss}A e_t + (I- \Phi(I + P_{ss}\Phi - A^\top)^{-1}  P_{ss}) \hat{w}.
\]
In the proof of Lemma~\ref{lem:rs_ss}, we have shown that $(I+\Phi P_{ss})^{-1}A$ is stable. Thuse, $\{ \tilde{x}_t\}$ converges to~\eqref{conv_state} as $t$ tends to infinity. Since $\bbar{x}_t = \tilde{x}_t - e_t$, $\{\bbar{x}_t\}$ also converges to~\eqref{conv_state}. 

Moreover, if $\hat{w}=0$, then
$\lim_{t\to\infty} \tilde{x}_t = 0$ and $\lim_{t\to\infty} \bbar{x}_t = 0$ as desired.
\end{proof}

\subsection{Proof of Proposition~\ref{prop:bibo}}\label{app:bibo}
\begin{proof}
Consider an adversarial policy $\gamma' \in \Gamma$ that maps the information vector to some distribution with a mean vector $\bar{w}_t$ and a covariance matrix $\Sigma$, such that the pair $(A, \Sigma^{1/2})$ is stabilizable. When the policy pair $(\pi_{ss}^*, \gamma')$ is applied to the mean-state system, the error state defined in the proof of Proposition~\ref{prop:stability}
has  the following form:
\[
e_{t+1} = (A - \bar{X}_{ss,\gamma'}^{-} C^\top (C \bar{X}_{ss,\gamma'}^{-} C^\top + M)^{-1} CA)e_t,
\]
where $\bar{X}_{ss,\gamma'}^{-}$ is the solution to the filter ARE~\eqref{cov_update_ss} with disturbance distribution $\mathbb{P}_t = \gamma'(I_t)$. Analogous to the proof of Proposition~\ref{prop:stability}, the error state $e_t$ converges to the origin regardless of the control gain matrix $K_{ss}$ since $(A - \bar{X}_{ss,\gamma'}^{-} C^\top (C \bar{X}_{ss,\gamma'}^{-} C^\top + M)^{-1} CA)$ has eigenvalues strictly within the unit circle. The expected value of the state estimate for the mean-state system can now be written as
\begin{equation}\label{mean_state_1}
\bbar{x}_{t+1} = \tilde{A} \bbar{x}_t + B L_{ss} + \mathbb{E}[{w}_t] + \bar{X}_{ss,\gamma'} C^\top M^{-1} C A e_t,
\end{equation}
where $\tilde{A}:=A+B K_{ss}$ is the closed-loop gain matrix and $\bar{X}_{ss,\gamma'}$ is the conditional state covariance matrix under the adversary's policy $\gamma'$. When viewing the disturbances $w_t$ as input, the above system is BIBO stable as long as $\mathbb{E}[w_t]$ is bounded and the matrix $\tilde{A}$ has eigenvalues strictly within the unit circle. Therefore, it is sufficient to show that for the system
\begin{equation}\label{closed_loop}
\bbar{x}_{t+1} = \tilde{A} \bbar{x}_t
\end{equation}
with an arbitrary initial state $\bbar{x}_0$, the expected value of the estimated state converges to the origin, i.e, $\bbar{x}_t\to 0$ as $t\to\infty$.

Using the closed-loop system  matrix $\tilde{A}$, the ARE~\eqref{are} is equivalent to
\begin{equation*}
P_{ss} = Q + \tilde{A}^\top P_{ss} \tilde{A} + K_{ss}^\top R K_{ss} + \tilde{A}^\top P_{ss} (\lambda I - P_{ss})^{-1} P_{ss} \tilde{A}.
\end{equation*}
Therefore, we have
\[
\begin{split}
\bbar{x}_{t+1}^\top  P_{ss} \bbar{x}_{t+1} - \bbar{x}_{t}^\top P_{ss} \bbar{x}_{t} & = \bbar{x}_t^\top (\tilde{A}^\top P_{ss} \tilde{A} - P_{ss}) \bbar{x}_t \\
& = - \bbar{x}_t^\top ( Q + K_{ss}^\top R K_{ss} + \tilde{A}^\top P_{ss}(\lambda I - P_{ss})^{-1} P_{ss} \tilde{A})\bbar{x}_t \\
& \leq 0,
\end{split}
\]
where the last inequality follows from $Q \succeq 0$, $R \succ 0$ and $(\lambda I - P_{ss})^{-1} \succ 0$ under Assumption~\ref{ass:lambda_ass}.
We also deduce that
\[
\bbar{x}_{t+1}^\top P_{ss} \bbar{x}_{t+1} = \bbar{x}_0^\top P \bbar{x}_0 - \sum_{k=0}^{t} \bbar{x}_k^\top ( Q + K_{ss}^\top R K_{ss} + \tilde{A}^\top P_{ss}(\lambda I - P_{ss})^{-1} P_{ss} \tilde{A}) \bbar{x}_k.
\]
However, as $P_{ss} \succeq 0$, the left-hand side of the above inequality is no less than zero. 
Since we have already shown that $\bbar{x}_t^\top ( Q + K_{ss}^\top R K_{ss} + \tilde{A}^\top P_{ss}(\lambda I - P_{ss})^{-1} P_{ss} \tilde{A})\bbar{x}_t \geq 0$ for each $t$,  
\[
\lim_{t\to\infty} \bbar{x}_t^\top (Q + K_{ss}^\top R K_{ss} + \tilde{A}^\top P_{ss}(\lambda I - P_{ss})^{-1} P_{ss} \tilde{A}) \bbar{x}_t = 0.
\]
This implies that
\begin{equation}\label{lims}
\lim_{t\to\infty} Q^{1/2} \bbar{x}_t = 0,\quad \lim_{t\to\infty} K_{ss}\bbar{x}_t = 0.
\end{equation}

Recall that $(A, Q^{1/2})$ is observable under Assumption~\ref{ass:stab_obs}. Furthermore, the relation $\bbar{x}_{t+1} = (A + B K_{ss})\bbar{x}_{t}$ yields
\[
\begin{bmatrix}
 Q^{1/2} (\bbar{x}_{t+n_x - 1} - \sum_{i=1}^{n_x - 1} A^{i-1} B K_{ss} \bbar{x}_{t + n_x - i - 1} )\\
Q^{1/2} (\bbar{x}_{t+n_x - 2} - \sum_{i=1}^{n_x - 2} A^{i-1} B K_{ss} \bbar{x}_{t + n_x - i - 2} )\\
\vdots\\
Q^{1/2} (\bbar{x}_{t+1} - B K_{ss} \bbar{x}_t)\\
Q^{1/2} \bbar{x}_t
\end{bmatrix} = \begin{bmatrix}
Q^{1/2} A^{n_x - 1} \\ Q^{1/2} A^{n_x - 2} \\ \vdots \\ Q^{1/2} A \\ Q^{1/2}
\end{bmatrix} \bbar{x}_t.
\]
From~\eqref{lims} the left-hand side tends to zero and hence the right-hand side also tends to zero. However, by the observability assumption the matrix on the right-hand side has full rank, implying that $\bbar{x}_t \to 0$. Therefore, the eigenvalues of $\tilde{A}$ lie strictly within the unit circle, and the system~\eqref{mean_state_1} is BIBO stable. Since $\tilde{x}_t = e_t - \bbar{x}_t$ and $\mathbb{E}[y_t] = C \tilde{x}_t$, we conclude that the mean-state system is also BIBO stable.
\end{proof}

\bibliographystyle{IEEEtran}

\bibliography{reference}

\begin{thebibliography}{10}
\providecommand{\url}[1]{#1}
\csname url@samestyle\endcsname
\providecommand{\newblock}{\relax}
\providecommand{\bibinfo}[2]{#2}
\providecommand{\BIBentrySTDinterwordspacing}{\spaceskip=0pt\relax}
\providecommand{\BIBentryALTinterwordstretchfactor}{4}
\providecommand{\BIBentryALTinterwordspacing}{\spaceskip=\fontdimen2\font plus
\BIBentryALTinterwordstretchfactor\fontdimen3\font minus
  \fontdimen4\font\relax}
\providecommand{\BIBforeignlanguage}[2]{{%
\expandafter\ifx\csname l@#1\endcsname\relax
\typeout{** WARNING: IEEEtran.bst: No hyphenation pattern has been}%
\typeout{** loaded for the language `#1'. Using the pattern for}%
\typeout{** the default language instead.}%
\else
\language=\csname l@#1\endcsname
\fi
#2}}
\providecommand{\BIBdecl}{\relax}
\BIBdecl

\bibitem{Hakobyan2022conf}
A.~Hakobyan and I.~Yang, ``Wasserstein distributionally robust control of
  partially observable linear systems: Tractable approximation and performance
  guarantee,'' in \emph{Proceedings of the 61st IEEE Conference on Decision and
  Control}, 2022.

\bibitem{aastrom2012introduction}
K.~J. {\AA}str{\"o}m, \emph{Introduction to Stochastic Control Theory}.\hskip
  1em plus 0.5em minus 0.4em\relax Courier Corporation, 2012.

\bibitem{khalil1996robust}
I.~Khalil, J.~Doyle, and K.~Glover, \emph{Robust and Optimal Control}.\hskip
  1em plus 0.5em minus 0.4em\relax Prentice Hall, 1996.

\bibitem{Kumar2015}
P.~R. Kumar and P.~Varaiya, \emph{Stochastic Systems: Estimation,
  Identification, and Adaptive Control}.\hskip 1em plus 0.5em minus 0.4em\relax
  SIAM, 2015.

\bibitem{Nilim2005}
A.~Nilim and L.~{El Ghaoui}, ``Robust control of {Markov} decision processes
  with uncertain transition matrices,'' \emph{Oper. Res.}, vol.~53, no.~5, pp.
  780--798, 2005.

\bibitem{Samuelson2017}
S.~Samuelson and I.~Yang, ``Data-driven distributionally robust control of
  energy storage to manage wind power fluctuations,'' in \emph{Proceedings of
  the 1st IEEE Conference on Control Technology and Applications}, 2017.

\bibitem{petersen2000minimax}
I.~R. Petersen, M.~R. James, and P.~Dupuis, ``Minimax optimal control of
  stochastic uncertain systems with relative entropy constraints,'' \emph{IEEE
  Transactions on Automatic Control}, vol.~45, no.~3, pp. 398--412, 2000.

\bibitem{ugrinovskii2002minimax}
V.~A. Ugrinovskii and I.~R. Petersen, ``Minimax {LQG} control of stochastic
  partially observed uncertain systems,'' \emph{SIAM Journal on Control and
  Optimization}, vol.~40, no.~4, pp. 1189--1226, 2002.

\bibitem{van2015distributionally}
B.~P. Van~Parys, D.~Kuhn, P.~J. Goulart, and M.~Morari, ``Distributionally
  robust control of constrained stochastic systems,'' \emph{IEEE Transactions
  on Automatic Control}, vol.~61, no.~2, pp. 430--442, 2015.

\bibitem{Yang2018}
I.~Yang, ``A dynamic game approach to distributionally robust safety
  specifications for stochastic systems,'' \emph{Automatica}, vol.~94, pp.
  94--101, 2018.

\bibitem{Tzortzis2019}
I.~Tzortzis, C.~D. Charalambous, and T.~Charalambous, ``Infinite horizon
  average cost dynamic programming subject to total variation distance
  ambiguity,'' \emph{SIAM J. Control Optim.}, vol.~57, no.~4, pp. 2843--2872,
  2019.

\bibitem{Coppens2021}
P.~Coppens and P.~Patrinos, ``Data-driven distributionally robust mpc for
  constrained stochastic systems,'' \emph{IEEE Control Systems Letters},
  vol.~6, no. 1274--1279, 2021.

\bibitem{schuurmans2021data}
M.~Schuurmans and P.~Patrinos, ``Data-driven distributionally robust control of
  partially observable jump linear systems,'' in \emph{Proceedings of the 60th
  IEEE Conference on Decision and Control}, 2021, pp. 4332--4337.

\bibitem{yang2020wasserstein}
I.~Yang, ``Wasserstein distributionally robust stochastic control: A
  data-driven approach,'' \emph{IEEE Transactions on Automatic Control},
  vol.~66, no.~8, pp. 3863--3870, 2021.

\bibitem{coulson2021distributionally}
J.~Coulson, J.~Lygeros, and F.~D{\"o}rfler, ``Distributionally robust chance
  constrained data-enabled predictive control,'' \emph{IEEE Transactions on
  Automatic Control}, 2021.

\bibitem{Mark2021}
C.~Mark and S.~Liu, ``Data-driven distributionally robust {MPC}: An indirect
  feedback approach,'' \emph{arXiv preprint arXiv:2109.09558}, 2021.

\bibitem{Tzortzis2021}
I.~Tzortzis, C.~D. Charalambous, and C.~N. Hadjicostis, ``A distributionally
  robust {LQR} for systems with multiple uncertain players,'' in
  \emph{Proceedings of the 60th IEEE Conference on Decision and Control}, 2021.

\bibitem{hakobyan2021wasserstein}
A.~Hakobyan and I.~Yang, ``Wasserstein distributionally robust motion control
  for collision avoidance using conditional value-at-risk,'' \emph{IEEE
  Transactions on Robotics}, vol.~38, no.~2, pp. 939--957, 2022.

\bibitem{Zolanvari2021}
A.~Zolanvari and A.~Cherukuri, ``Data-driven distributionally robust iterative
  risk-constrained model predictive control,'' in \emph{Proceedings of 2022
  European Control Conference}, 2022.

\bibitem{kim2022minimax}
K.~Kim and I.~Yang, ``Distributional robustness in minimax linear quadratic
  control with {W}asserstein distance,'' \emph{SIAM Journal on Control and
  Optimization}, 2022.

\bibitem{Zhong2022}
Z.~Zhong, E.~A. del Rio-Chanona, and P.~Petsagkourakis, ``Distributionally
  robust {MPC} for nonlinear systems,'' in \emph{IFAC-PapersOnLine}, vol.~55,
  no.~7, 2022, pp. 606--613.

\bibitem{Dixit2022}
A.~Dixit, M.~Ahmadi, and J.~W. Burdick, ``Distributionally robust model
  predictive control with total variation distance,'' \emph{arXiv preprint
  arXiv:2203.12062}, 2022.

\bibitem{Micheli2022}
F.~Micheli, T.~Summers, and J.~Lygeros, ``Data-driven distributionally robust
  {MPC} for systems with uncertain dynamics,'' \emph{arXiv preprint
  arXiv:2209.08869}, 2022.

\bibitem{calafiore2007ambiguous}
G.~C. Calafiore, ``Ambiguous risk measures and optimal robust portfolios,''
  \emph{SIAM Journal on Optimization}, vol.~18, no.~3, pp. 853--877, 2007.

\bibitem{delage2010distributionally}
E.~Delage and Y.~Ye, ``Distributionally robust optimization under moment
  uncertainty with application to data-driven problems,'' \emph{Operations
  Research}, vol.~58, no.~3, pp. 595--612, 2010.

\bibitem{wiesemann2014distributionally}
W.~Wiesemann, D.~Kuhn, and M.~Sim, ``Distributionally robust convex
  optimization,'' \emph{Operations Research}, vol.~62, no.~6, pp. 1358--1376,
  2014.

\bibitem{ben2013robust}
A.~Ben-Tal, D.~Den~Hertog, A.~De~Waegenaere, B.~Melenberg, and G.~Rennen,
  ``Robust solutions of optimization problems affected by uncertain
  probabilities,'' \emph{Management Science}, vol.~59, no.~2, pp. 341--357,
  2013.

\bibitem{bayraksan2015data}
G.~Bayraksan and D.~K. Love, ``Data-driven stochastic programming using
  phi-divergences,'' in \emph{The Operations Research Revolution}.\hskip 1em
  plus 0.5em minus 0.4em\relax INFORMS, 2015, pp. 1--19.

\bibitem{mohajerin2018data}
P.~Mohajerin~Esfahani and D.~Kuhn, ``Data-driven distributionally robust
  optimization using the {W}asserstein metric: Performance guarantees and
  tractable reformulations,'' \emph{Mathematical Programming}, vol. 171, no.~1,
  pp. 115--166, 2018.

\bibitem{gao2016distributionally}
R.~Gao and A.~J. Kleywegt, ``Distributionally robust stochastic optimization
  with {W}asserstein distance,'' \emph{arXiv preprint arXiv:1604.02199}, 2016.

\bibitem{zhao2018data}
C.~Zhao and Y.~Guan, ``Data-driven risk-averse stochastic optimization with
  {W}asserstein metric,'' \emph{Operations Research Letters}, vol.~46, no.~2,
  pp. 262--267, 2018.

\bibitem{kuhn2019wasserstein}
D.~Kuhn, P.~M. Esfahani, V.~A. Nguyen, and S.~Shafieezadeh-Abadeh,
  ``Wasserstein distributionally robust optimization: Theory and applications
  in machine learning,'' in \emph{Operations Research \& Management Science in
  the Age of Analytics}.\hskip 1em plus 0.5em minus 0.4em\relax INFORMS, 2019,
  pp. 130--166.

\bibitem{coppens2020data}
P.~Coppens, M.~Schuurmans, and P.~Patrinos, ``Data-driven distributionally
  robust {LQR} with multiplicative noise,'' in \emph{Learning for Dynamics and
  Control}.\hskip 1em plus 0.5em minus 0.4em\relax PMLR, 2020, pp. 521--530.

\bibitem{Mark2020}
C.~Mark and S.~Liu, ``Stochastic {MPC} with distributionally robust chance
  constraints,'' in \emph{Proceedings of the 21st IFAC World Congress}, 2020.

\bibitem{Boskos2020}
D.~Boskos, J.~Cort\'{e}s, and S.~Mart\'{i}nez, ``Data-driven ambiguity sets
  with probabilistic guarantees for dynamic processes,'' \emph{IEEE
  Transactions on Automatic Control}, 2020.

\bibitem{ugrinovskii1999finite}
V.~A. Ugrinovskii and I.~R. Petersen, ``Finite horizon minimax optimal control
  of stochastic partially observed time varying uncertain systems,''
  \emph{Mathematics of Control, Signals and Systems}, vol.~12, no.~1, pp.
  1--23, 1999.

\bibitem{nakao2021distributionally}
H.~Nakao, R.~Jiang, and S.~Shen, ``Distributionally robust partially observable
  {M}arkov decision process with moment-based ambiguity,'' \emph{SIAM Journal
  on Optimization}, vol.~31, no.~1, pp. 461--488, 2021.

\bibitem{Hakobyan2021map}
A.~Hakobyan and I.~Yang, ``Distributionally robust risk map for learning-based
  motion planning and control: A semidefinite programming approach,''
  \emph{IEEE Transactions on Robotics}, 2022.

\bibitem{osogami2015robust}
T.~Osogami, ``Robust partially observable {M}arkov decision process,'' in
  \emph{International Conference on Machine Learning}.\hskip 1em plus 0.5em
  minus 0.4em\relax PMLR, 2015, pp. 106--115.

\bibitem{saghafian2018ambiguous}
S.~Saghafian, ``Ambiguous partially observable {M}arkov decision processes:
  Structural results and applications,'' \emph{Journal of Economic Theory},
  vol. 178, pp. 1--35, 2018.

\bibitem{Gonzalez2003}
J.~I. Gonz\'{a}lez-Trejo, O.~{Hern\'{a}ndez-Lerma}, and L.~F. Hoyos-Reyes,
  ``Minimax control of discrete-time stochastic systems,'' \emph{SIAM J.
  Control Optim.}, vol.~41, no.~5, pp. 1626--1659, 2003.

\bibitem{Hernandez2012}
O.~{Hern\'{a}ndez-Lerma} and J.~B. Lasserre, \emph{Discrete-Time Markov Control
  Processes: Basic Optimality Criteria}.\hskip 1em plus 0.5em minus 0.4em\relax
  Springer, 2012.

\bibitem{anderson2012optimal}
B.~D. Anderson and J.~B. Moore, \emph{Optimal Filtering}.\hskip 1em plus 0.5em
  minus 0.4em\relax Courier Corporation, 2012.

\bibitem{o2016conic}
B.~O'donoghue, E.~Chu, N.~Parikh, and S.~Boyd, ``Conic optimization via
  operator splitting and homogeneous self-dual embedding,'' \emph{Journal of
  Optimization Theory and Applications}, vol. 169, no.~3, pp. 1042--1068, 2016.

\bibitem{andersen2003implementing}
E.~D. Andersen, C.~Roos, and T.~Terlaky, ``On implementing a primal-dual
  interior-point method for conic quadratic optimization,'' \emph{Mathematical
  Programming}, vol.~95, no.~2, pp. 249--277, 2003.

\bibitem{aps2019mosek}
M.~ApS, ``{MOSEK} optimization suite,'' 2019.

\bibitem{kailath2000linear}
T.~Kailath, A.~H. Sayed, and B.~Hassibi, \emph{Linear Estimation}.\hskip 1em
  plus 0.5em minus 0.4em\relax Prentice Hall, 2000.

\bibitem{Fournier2015}
N.~Fournier and A.~Guillin, ``On the rate of convergence in {Wasserstein}
  distance of the empirical measure,'' \emph{Probability Theory and Related
  Fields}, vol. 162, no. 3--4, pp. 707--738, 2015.

\bibitem{bertsekas2012dynamic}
D.~Bertsekas, \emph{Dynamic Programming and Optimal Control: Volume I}.\hskip
  1em plus 0.5em minus 0.4em\relax Athena Scientific, 2012, vol.~1.

\end{thebibliography}

\end{document}